\documentclass[preprint,review,11pt]{elsarticle}
\usepackage[T1]{fontenc}        %arruma hifenação com palavras acentuadas
\usepackage[utf8]{inputenc}
\usepackage[verbose,a4paper,left=20mm,right=20mm,top=30mm,bottom=30mm]{geometry}
\usepackage{color,soul}

\usepackage[normalem]{ulem}
\usepackage{url}
\usepackage{bbm}
\usepackage{textcomp}
\usepackage{epsfig}
\usepackage[normalem]{ulem}
\usepackage{acronym}
\usepackage{fancyhdr}
\usepackage[ruled,vlined,linesnumbered]{algorithm2e}
\usepackage{colortbl}
\usepackage{longtable}
\usepackage{amsmath,amsfonts,float}
\usepackage{graphicx}
\usepackage{amssymb}
\usepackage{amsthm}
\usepackage{lineno}
\usepackage{natbib}
\usepackage{tikz} \usetikzlibrary{shapes,arrows,fit,calc,plotmarks,positioning} 
\usepackage{pgfplots}
\pgfplotsset{width=9cm,compat=1.9}
 \usepgfplotslibrary{statistics}
\usepackage{multirow}
\usepackage{verbatim}
\usepackage{caption}
\usepackage{subfigure}
\usepackage{wrapfig}
\usepackage{tablefootnote}
\usepackage{threeparttable}
\usepackage{setspace}

\pgfplotscreateplotcyclelist{mylist}{
{black,mark=+},
{blue,mark=+},
{magenta,mark=+},
{orange,mark=+}% <-- don’t add a comma here
}

\date{}

\begin{document}
\onehalfspacing

\begin{frontmatter}

\title{Hybrid matheuristics to solve the integrated lot sizing and scheduling problem on parallel machines with sequence-dependent and non-triangular setup} 

\author[label1]{Desiree M.~Carvalho\corref{cor1}}
\ead{dmcarvalho@unifesp.br}
\cortext[cor1]{Corresponding author}

\author[label1]{Mari\'{a} C.~V.~Nascimento}
\ead{mcv.nascimento@unifesp.br}

\address[label1]{Instituto de Ci\^{e}ncia e Tecnologia, Universidade Federal de S\~{a}o Paulo - UNIFESP\\ 
Av. Cesare G. Lattes, 1201, S\~{a}o Jos\'{e} dos Campos, SP, Brasil \\}

%%%%%%%%%%%%%%%%%%%%%%%%%%%%%%%%%%%%%%%%%%%%%%%%%%%%%%%%%%%%%%%%%%%%

\begin{abstract}
This paper approaches the integrated lot sizing and scheduling problem (ILSSP), in which non-identical machines work in parallel with non-triangular sequence-dependent setup costs and times, setup carry-over and capacity limitation. The aim of the studied ILSSP, here called ILSSP-NT on parallel machines, is to determine a production planning and tasks sequencing that meet period demands without delay and in such a way that the total costs of production, machine setup and inventory are minimized. The dearth of literature on the ILSSP-NT, despite the increasing amount of applications in the industrial sector, mainly in the food processing industry, motivated us to conduct this study. In this paper, we propose efficient methods to solve the ILSSP-NT on parallel machines. The methods virtually consist in the hybridization of the relax-and-fix and fix-and-optimize methods with the path-relinking and kernel search heuristics. To assess how well the heuristics solve the ILSSP-NT on parallel machines, we compared their results with those of the CPLEX solver with a fixed CPU time limit. The proposed matheuristics significantly outperformed CPLEX in most of the tested instances.
\end{abstract}


\end{frontmatter}

%\linenumbers

\section{Introduction}
\label{sec:intro}

In the industrial sector, production planning leans on efficient decision-making strategies in order for the company to gain an edge on the competitive market. Production planning aims at optimizing the resources of a company to meet its long, medium and short-term objectives. Roughly speaking, long-term decisions belong to the strategic level of production planning, when a company's global goals are defined. Tactical planning is responsible for efficiently using the available resources in order to fulfill the goals of the strategic planning. In the short-term planning level, known as operational planning, decisions related to the company's routine are made.

Production planning for a variety of realistic scenarios involves deciding the number of lots of items to be produced (lot sizing), and the planning and sequencing of tasks to be performed for the production of the items (scheduling) \citep{Araujo2007,Marinelli2007,Ferreira2009,Boonmee2016}. One may define a production plan by considering lot sizing and scheduling separately, where the lot sizing problem is solved first so that the tasks can be scheduled. This approach is suitable for problems whose costs and machine setup time do not depend on the sequence of tasks performed during the production of the items. Nevertheless, when there is a dependence between machine setup and sequence of tasks, e.g, when capacity use violates an imposed limit, the results of the scheduling also affect the lot sizing. Therefore, it is not possible to approach lot sizing and scheduling in a hierarchical and separate manner.

Better decision-making strategies to solve problems with sequence-dependent setup costs and times may be reached by simultaneously integrating the production plan considering lot sizing and scheduling (integrated) \citep{Guimaraes2013,Guimaraes14}. The resulting problem is known as integrated lot sizing and scheduling problem (ILSSP), which is recognized as being computationally challenging. Classic ILSSPs aim at defining a production plan that minimizes the total costs of production in a finite planning horizon. The ILSSP is an extension of the single item capacitated lot sizing problem (CLSP) with setup costs. Therefore, the problem of determining an optimal solution to the ILSSP is $\mathcal{NP}$-hard \citep{FL80}. Moreover, the feasibility problem of ILSSP is $\mathcal{NP}$-complete \citep{MM91}.

Heuristic methods efficiently solve several variants of the ILSSP, even the significantly hard ins\-tances. Furthermore, the advances in computational power predicted by Moore's law enable the incorporation of characteristics in ILSSP models that better approximate the mathematical formulations to real scenarios. Considering these facts, we propose solution methods to efficiently solve a relevant problem that has not yet been extensively investigated by the literature, the ILSSP on parallel machines with non-triangular sequence-dependent setup\footnote{Producing an item $j$ immediately after an item $i$ may not be the best environmental option, since the production of an item $l$ that absorbs contamination as intermediary of $i$ and $j$ would diminish the costs and times of cleansing operations. In other words, considering that $s_{ij}$ is the setup cost/time necessary to stop the production of item $i$ and to start the production of item $j$, the triangular inequality $s_{il} + s_{lj} \geq s_{ij}$ does not hold and $l$ is said to be a shortcut item.} costs and times and setup carry-over, referred to here as ILSSP-NT on parallel machines.

Important applications of the ILSSP-NT on parallel machines can be found in the food processing industry \citep{Claassen2016,Toso2009}, e.g., in the animal feed industry. In the last five years the production of animal feed increased by an average of 2.49\% a year \citep{Alltech2018} and, according to the IFIF (International Feed Industry Federation), it tends to soar until 2050 \citep{ifif}. The growth in the animal feed industry is one of the consequences of the rise in consumption of raw materials such as meat, milk and eggs provided by the livestock industry.

To the best of our knowledge, a few solution methods to approach the ILSSP-NT on parallel machines can be found in the literature. \citet{Kang1999} proposed a branch-and-price-based approach to solve the ILSSP-NT on parallel machines without setup times.  \citet{Meyr02} introduced four heuristics, which are combinations of the threshold acceptance and simulated annealing heuristics with dual re-optimization. On the other hand, many problems related to the ILSSP-NT on parallel machines have been investigated \citep{Toso2009,Menezes11,James2011,Guimaraes2013,Guimaraes14,clark2014,Xiao2015, Claassen2016,Martinez2018}.

Mathematical programming-based heuristics, also known as matheuristics, e.g., the relax-and-fix and fix-and-optimize heuristics, have been widely suggested to approach problems related to the ILSSP-NT \citep{Toso2009,James2011,Guimaraes2013}. Moreover, the results presented in \citep{James2011} indicate that local search-based methods play an important role in improving the quality of the solutions. Therefore, in this paper we propose solution methods that combine the relax-and-fix and fix-and-optimize heuristics and that use intensification-based heuristics known as path-relinking \citep{glover1997} and kernel search \citep{angelelli2010} to refine the solutions. The introduced intensification strategies were designed to consider the particularities of the non-triangular setups to better approach the ILSSP-NT.

Computational experiments were carried out with the proposed heuristics using instance sets obtained from studies in the literature. In the first experiment, we employed the introduced solution methods to approach the ILSSP-NT on a single machine on the instances introduced in \citep{Guimaraes2013}. The results obtained by our solution methods are contrasted to those found by the CPLEX v.  12.10 solver with an imposed time limit of 3600 seconds. In this experiment, the proposed heuristics were very competitive with CPLEX, achieving better average gaps in the problems tested. In the second experiment, we tested the heuristics presented here to address the ILSSP-NT on parallel machines. In this experiment, we used instances proposed in \citep{James2011} but modified the setup costs and times of some items so that they were shortcut items, as introduced in \citep{Guimaraes2013}, and compared the solutions obtained using the proposed heuristics with the solutions found by the CPLEX v. 12.10 software \citep{cplex2017} time limited to 3600 seconds. The results of this experiment demonstrate the outstanding performance of the proposed matheuristics with the intensification strategies in comparison to CPLEX.

This paper is organized as follows: Section \ref{pidls} presents a review of problems related to the ILSSP; Section \ref{sec:problema} presents the main characteristics of the ILSSP-NT on parallel machines as well as the investigated mathematical formulation; Section \ref{sec:RFO} provides a thorough description of all the methods proposed to solve the ILSSP-NT on parallel machines; Section \ref{TestPre} presents a comparative analysis of the results obtained by the proposed heuristics and the CPLEX v. 12.10 solver when solving the ILSSP-NT on a single and parallel machines; and finally, Section \ref{conclu} presents the final remarks and directions for future research.

\section{Integrated lot sizing and scheduling problem (ILSSP)}
\label{pidls}

In the context of this paper, for a finite planning horizon, the lot sizing problem aims at defining how much and when to produce items in a production line to satisfy their pre-established demands respecting the available resources and the machine capacity limit for each period of the planning horizon. The scheduling problem, on the other hand, makes sure that there are no conflicts of items sharing resources in the production process, e.g., the machines. For this, the scheduling problem defines the sequencing of operations that must be performed, such as machine setup and items production \citep{DK97}.

The two problems can be considered separately when preparing the company planning \citep{KG03,Busch2010}. Nevertheless, integrating them into a single problem may provide a better production planning. In this sense, the goals of the ILSSP may vary. For example, a commonly studied variant of the ILSSP aims to minimize the production makespan, while another seeks to minimize the total costs involved in the production process. In both cases, demands for the items must be met, restrict to the available resources. Some literature reviews on the ILSSP can be found in the literature \citep{DK97,Zhu2006,Jans08,Clark11,Guimaraes14,Copil2017}.

\citet{Zhu2006} focused their review on lot sizing and scheduling problems with sequence-dependent setup times and costs. Problems with different types of machine configuration were investigated, including those involving single machine, parallel machines, flow shop, and job shop systems. The authors highlight the relevance of studying these problems and draw attention to the lack of solution methods capable of efficiently solving them despite the considerable amount of attention they have received.

In \citep{Clark11}, the authors emphasize the importance of integrating both the lot sizing and the scheduling problems, as well as considering the integration of other problems such as distribution and vehicle routing. In line with this, \citet{Clark11} point out the need for efficient solution methods and tight formulations mainly in cases where models incorporate many elements to describe integrated problems more precisely. In particular, to effectively solve these problems, the authors strongly suggest solution methods that are hybrids of exact and heuristic methods.

The reviews on the ILSSP published in the last decade attempt to present models and solution methods to case studies found in the industry. \citet{Jans08} expose several extensions of the problems presented in \citep{DK97} to meet the needs of the industries. Moreover, the authors suggest that studying the lot sizing problem on parallel machines and more general models are interesting subjects for future works. 

Among the methods proposed to solve the ILSSP on parallel machines, \citep{Kang1999,Meyr02,James2011} deserve to be highlighted. \citet{Kang1999} approached a number of business problems provided by private companies, known as CHES (Chesapeake Decision Sciences) \citep{Baker1989}. They considered CHES to be an extension of the ILSSP on parallel machines with sequence-dependent setup costs, whose triangular inequality may or may not hold, and null setup times. The model used by the authors considers the splitting of the production sequence of each period into smaller split-sequences, as they call them, and restrict the number of items to be produced per split-sequence. To solve the problems, they apply a branch-and-bound/column generation-based approach with two heuristics they proposed. One of the proposed heuristics truncates the branch-and-bound tree according to the probability of achieving a good solution by branching the current node, the other is an improvement heuristic that uses a local search method to find the best solutions in the neighborhood of the current solution. The results obtained by the method were satisfactory for instances with up to 95\% of machine capacity utilization. However, the method required a high computational time.

Study \citep{Meyr02} investigates the generalized lot sizing and scheduling problem on non-identical parallel machines with sequence-dependent setup costs and times that may or may not obey the triangular inequality. To solve the problem, the author proposes heuristics that combine the local search methods threshold accepting \citep{Dueck1990} and simulated annealing \citep{Kirkpatrick1983} with dual optimization. In the first step, local search methods are responsible for fixing setup patterns. Then, a solution is obtained for each fixed pattern by solving the dual re-optimization of a network flow problem using the algorithm proposed by \citet{Bertsekas1988}. The author observes that despite the competitive results with regard to those obtained in \citep{Kang1999}, the computational times of the proposed methods were very high.

To the best of our knowledge, the most state-of-the-art solution methods to solve the ILSSP on parallel machines were proposed in \citep{James2011}. Several iterative MIP-based neighborhood search heuristics are introduced in \citep{James2011}. The main strategy is a stochastic MIP-based local search heuristic that consists of three steps. In the first step, an initial solution is obtained through a general relax-and-fix framework. In the second step, the solution is improved by an intensification heuristic (local search), identified by the authors as XPH. Several subMIPs are randomly selected by the intensification algorithm and solved through a fix-and-optimize heuristic \citep{Pochet2006}. The probability of a subMIP being selected is related to its frequency and time interval, which starts when the subMIP is last selected. In this sense, the higher the frequency and the shorter the referred time interval, the smaller the chances of this subMIP being selected again. Finally, the third step is responsible for the diversification of the solution method so that it does not get trapped in a local minimum through a Variable Neighborhood Search (VNS) \citep{Almada2010}.

\citet{James2011} proposed the INSRF heuristic to solve the ILSSP on parallel machines whose initial solution is found by a relax-and-fix heuristic. They compared INSRF with other methods, among which XPHRF, FOHRF9 and CPLEX. XPHRF combines the stochastic MIP-based local search XPH with a constructive relax-and-fix heuristic. The FOHRF9 heuristic is the combination of the constructive heuristic relax-and-fix with the fix-and-optimize heuristic from \citet{Sahling2009}. With a time limit of 3600 seconds to solve each problem, the best results were obtained by the XPHRF heuristic.

A practical application of ILSSP-NT on a single machine to heuristically solve the production planning of an animal feed factory is presented by \citet{Toso2009}. Due to the need for cleansing processes in the production process, the authors consider sequence-dependent setup times that do not obey the triangular inequality. The three solution methods they propose are based on the relax-and-fix heuristic. Experiments using real data demonstrated that the proposed modeling and methods significantly improved factory planning. Later, \citet{Clark2010} approached the same problem and proposed two models to address the ILSSP-NT based on the asymmetric traveling salesman problem (ATSP), where one considers setup carry-over and the other does not. Moreover, to solve the problems, the authors developed two methods: (i) an ATSP subtour elimination method; and (ii) an ATSP subtour elimination with a patching heuristic \citep{karp1979}.

\citet{Guimaraes2013} also investigated the ILSSP-NT on a single machine. The authors heuristically solved the ILSSP-NT through a strategy that is a hybrid of a column generation method and a relax-and-fix heuristic to find an initial feasible solution. This method is followed by an improvement heuristic that combines a column generation method with a fix-and-optimize heuristic. The results of the experiments show the superiority of the proposed method over the methods proposed by \citet{Kang1999} and \citet{Meyr02} when solving the small-sized instances with 6 items and 9 periods proposed in \citep{Kang1999}.  Experiments with the proposed methods to approach the ILSSP on a single machine attest the superiority of their results to those presented in \citep{James2011}, in particular, for larger instances.

According to \citet{Guimaraes14}, in the last years, commercial solvers have satisfactorily solved the ILSSP not considering setup costs and sequence-dependent setup times. This was achieved in reasonable times. To deal with the subject, \citet{Guimaraes14} focused on analysing the performance of commercial solvers on four formulations proposed in the literature that approach ILSSPs with sequence-dependent setup costs and times.

The analysis of the models conducted by \citet{Guimaraes14} also took into account the performance of a commercial solver considering different formulations of the ILSSP- NT. In addition, the authors introduced a new model to the ILSSP-NT and a facility location problem reformulation suggested by \citet{KB77}. Computational experiments indicated that, on average, the solver achieved the best performance when considering the introduced reformulation. Bearing this in mind, we opted to adapt such formulation to tackle the ILSSP-NT on parallel machines. The next section describes the problem under study and the suggested mathematical model. 

\section{Problem formulation}
\label{sec:problema}

This study approaches a single stage production planning problem with multiple items on non-identical parallel machines. The planning horizon is finite and subdivided into macro-periods. The machines have a pre-determined time limited production capacity and item demands are provided per period. In addition, to produce the demands of a given item, the machine setup must be performed. In particular, if a machine is ready to produce a certain item at the beginning of a period as it was the last item to be produced in previous periods, setup carry-over can be considered in order to avoid redundant machine setups. Therefore, one setup carry-over is allowed per pair of adjacent periods. 

In order to diminish the number of machine cleansing operations, non-triangular setups can be considered when items have the property of absorbing contamination during their production process. Due to non-triangular setup costs and times, a minimum lot size is considered to avoid fictitious setups associated to the production of empty lots. Furthermore, the production of multiple lots of an item is possible in the same period. The production plan must respect the production capacity to meet the item demands per period and to guarantee that operation conflicts do not happen during the production process. The goal of the problem is to find a production plan that meets all constraints by minimizing the setup and inventory costs.  

The mixed integer programming model used to solve the ILSSP-NT on parallel machines is based on the facility location problem \citep{KB77}. In the model, a feasible setup sequence is provided through a network flow as proposed in \cite{Guimaraes14}, where the node (vertex) of the network represents the production lot of an item and the arc connecting two nodes indicates the change in the production from an item to another. Before presenting the model, consider the following indexes, parameters and decision variables.

\begin{center}
\begin{longtable}{ccp{10cm}}
  Dimension parameters & & \\
   $n$ & : & number of items; \\
   $m$ & : & number of machines; \\
   $p$ & : & number of periods; \\
 Indexes & & \\
  $i,j$ & : & indexes representing items, $i,j \in \{1,\ldots,n\}$;  \\
  $k$ & : & index representing the machines, $k \in \{1,...,m\}$;  \\
  $t,u$ & : & indexes representing the macro-periods, $t,u \in \{1,...,p\}$; \\
  \noindent  Parameters & & \\
  $c_{ijk}$ & : & setup cost to change the state of machine $k$ from item $i$ to item $j$; \\
  $h_{i}$ & : & unitary inventory cost of item $i$; \\
  $d_{it}$ & : & demand of item $i$ in period $t$; \\
  $b_{ijk}$ & : & setup time to change the state of machine $k$ from item $i$ to item $j$; \\
  $M_{it}$ & : & maximum amount of item $i$ that can be produced in period $t$; \\
  $f_{ikt}$ & : & processing time of item $i$ at machine $k$ in period $t$; \\
  $m_{i}$ & : & minimum production lot size of item $i$; \\
  $T_{kt}$ & : & production capacity of machine $k$ in period $t$;  \\
  $q_{ikt}$ & : & maximum number of times machine $k$ can be set up to produce item $i$ in period $t$; \\
  Decision variables &  & \\
  $x_{iktu}$ & : & amount of item $i$ produced at machine $k$ in period $t$ to meet the demand of period $u$; \\
  $x^{b}_{ikt}$ & : & amount of item $i$ produced at machine $k$ in the beginning of period $t$, before the first machine setup is performed in period $t$; \\ 
  $x^{a}_{ikt}$ & : & amount of item $i$ produced at machine $k$ during period $t$, after the first machine setup is performed; \\ 
  $z_{ikt}$ & : & binary variable that assumes value 1 if machine $k$ is ready to produce item $i$ at the beginning of period $t$ (setup carry-over) and 0, otherwise;\\
  $y_{ijkt}$ & : & number of times a production process changes from item $i$ to item $j$ at machine $k$ in period $t$;\\
  $R_{kt}$ & : & binary variable that receives value 1 if at least one setup is performed at machine $k$ in period $t$\textcolor{red}{, i.e., ($\mathop {\sum }\limits_{i=1}^{n} \mathop {\sum } \limits_{j=1}^{n}y_{ijkt} \geq 1$),}  and 0, otherwise; \\
  $G_{ikt}$ & : & binary variable that receives value 1 if machine $k$ is ready at least once to produce item $i$ during period $t$ and 0, otherwise; \\
  $F_{ijkt}$ & : & commodity flow from node (item) $i$ to node (item) $j$ at machine $k$ in period $t$. \\
\end{longtable}
\end{center}

The ILSSP-NT on parallel machines is modeled as follows: 

\begin{align}
    \min  \mathop {\sum }\limits_{i=1}^{n} \mathop {\sum }\limits_{k=1}^{m} \mathop {\sum }\limits_{t=1}^{p}\mathop {\sum }\limits_{u=t}^{p}(u-t)h_{i}x_{iktu} + \mathop {\sum }\limits_{i=1}^{n} \mathop {\sum }
    \limits_{j=1}^{n}\mathop {\sum }\limits_{k=1}^{m}\mathop {\sum }\limits_{t=1}^{p}c_{ijk}y_{ijkt} & \label{nt:fo}
\end{align}

subject to:

    \begin{align}
    & \mathop {\sum }\limits_{k} \mathop {\sum }\limits_{t=1}^{u}x_{iktu} = d_{iu} & \forall & \; (i,u) & \label{eqmod1}\\
    &\mathop {\sum }\limits_{i}\left(\mathop {\sum }\limits_{u=t}^{p}f_{ikt}x_{iktu} + \mathop {\sum }\limits_{j}b_{jik}y_{jikt}\right) \leq T_{kt} & \forall & \; (k,t) & \label{eqmod2}\\
    &x_{iktu} \leq M_{it}G_{ikt} & \forall & \; (i,k,t,u) &\label{eqmod3} \\
    &\mathop {\sum }\limits_{i}z_{ikt} = 1 & \forall & \; (k,t) & \label{eqmod4} \\
    &z_{ikt} + \mathop {\sum }\limits_{j} y_{jikt} = \mathop {\sum }\limits_{j}y_{ijkt} + z_{ik,t+1}  & \forall & \; (i,k,t) &\label{eqmod5} \\
    &z_{ikt} + \mathop {\sum }\limits_{j} y_{jikt} \geq G_{ikt}  & \forall & \; (i,k,t) & \label{eqmod6} \\
	&z_{ikt} + \mathop {\sum }\limits_{j} y_{jikt} \leq q_{ikt}G_{ikt}  & \forall & \; (i,k,t) &\label{eqmod7} \\
	&\mathop {\sum }\limits_{u=t}^{p} x_{iktu} = x^{a}_{ikt} + x^{b}_{ikt} & \forall & \; (i,k,t) &\label{eqmod12} \\
	&x^{b}_{ikt} \leq M_{it} z_{ikt} & \forall & \; (i,k,t) &\label{eqmod13} \\
	&x^{a}_{ikt} \geq m_{i}\left(\mathop {\sum }\limits_{j} y_{jikt} - z_{ik,t+1} \right) & \forall & \; (i,j,k,t) &\label{eqmod14} \\
	&x^{a}_{ikt} + \mathop {\sum }\limits_{\lambda=t+1}^{u} x^{b}_{ik\lambda}  \geq m_{i}\mathop {\sum }\limits_{j} y_{jikt} - M_{it}\left(\mathop {\sum }\limits^{u-1}_{\lambda=t+1} R_{k\lambda} + 1 - R_{ku} \right) & \forall & \; (i,k,t,u), u \neq t &\label{eqmod15} \\
	& \mathop {\sum }\limits_{j}F_{0jkt} = \mathop {\sum }\limits_{i}G_{ikt} & \forall & \; (k,t) & \label{eqmod8} \\
    &F_{0ikt}+\mathop {\sum }\limits_{j}F_{jikt} = G_{ikt} + \mathop {\sum }\limits_{j}F_{ijkt} & \forall & \; (i,k,t) & \label{eqmod9} \\
    &F_{0ikt} \leq n z_{ikt} & \forall & \; (i,k,t) &\label{eqmod10} \\
	&F_{ijkt} \leq n y_{ijkt} & \forall & \; (i,j,k,t) &\label{eqmod11}\\
	&F_{ijkt} \geq 0  & \forall & \; (i,j,k,t) &\label{eqmod16} \\
	&G_{ikt}, R_{kt} \in \{0,1\}  & \forall & \; (i,k,t) &\label{eqmod17} \\
	&z_{ikt} \in \{0,1\}, x_{iktu} \geq 0  & \forall & \; (i,k,t,u) & \label{eqmod18} \\
	&y_{ijkt} \in \{0, \ldots, q_{jkt}\}  & \forall & \; (i,j,k,t) & \label{eqmod19}
    \end{align}
    
Objective function \eqref{nt:fo} expresses the sum of the machine setup and inventory costs to be minimized. Constraints \eqref{eqmod1}-\eqref{eqmod3} are the lot sizing constraints. Constraints \eqref{eqmod1} represent the inventory balance equations for each item and period. Constraints \eqref{eqmod2} limit the total production and setup times required by each machine and period. Constraints \eqref{eqmod3} guarantee that if an item is produced, its respective machine setup is considered.

Constraints \eqref{eqmod4}-\eqref{eqmod7} are responsible for keeping track of the setup state sequence. Constraints \eqref{eqmod4} restrict the machine setup to a single item at the beginning of each period. Constraints \eqref{eqmod5} ensure that the number of times the machine setup was ready to an item $i$ in period $t$ is equal to the number of times that there was a setup from the machine state $i$ to another setup state $j$ plus the possibility of the setup of item $i$ to be carried over the next period. Constraints \eqref{eqmod6} guarantee that if the machine was ready to produce an item $i$ in period $t$, then the correspondent setup state was carried over into period $t$ or at least one setup state changeover from an item $j$ to another item $i$ occurred in period $t$. Constraints \eqref{eqmod7} limit the number of times that machine $k$ can be ready to produce item $i$ during period $t$.

The production of a minimum lot size is enforced by Constraints \eqref{eqmod12}-\eqref{eqmod15} whenever a setup is performed. Constraints \eqref{eqmod12} split the total production of item $i$ at machine $k$ in period $t$ into the amount of item $i$ produced before the first machine setup is performed in period $t$ and the amount produced after the first setup up to the end of period $t$.  Constraints \eqref{eqmod13} impose that the production of item $i$ before the first machine setup in period $t$ only happens if the machine setup to item $i$ was carried over into period $t$. Constraints \eqref{eqmod14} ensure that a minimum amount of item $i$ is produced in period $t$ when the machine is set up for its production in period $t$ and this setup is not carried over into the next period. \textcolor{red}{Constraints \eqref{eqmod15} ensure that a minimum amount of item $i$ is produced in the following situations: (i) machine $k$ is set up for its production in period $t$; (ii) the production of item $i$ at machine $k$ starts in period $t$ and is carried over to the next periods. In the latter case, the production lot of item $i$ in period $t$ can be split into consecutive periods $t+1, \ldots, u$. The split is possible if no setup occurs in periods $t+1, \ldots, u$, which is identified through the auxiliary decision variables $R_{k\lambda}$, where $\lambda=\{t+1,...,u\}$.}

Constraints \eqref{eqmod8}-\eqref{eqmod11} are applied to guarantee that no sub-tour is created in the production sequencing. Constraints \eqref{eqmod8} set the origin (source) of the commodity flow in each period. Constraints \eqref{eqmod9} ensure the flow balance by sending a unitary flow to each selected node, generating single paths between every pair of items that begin at the origin and end at an item produced in that period. Constraints \eqref{eqmod10} and \eqref{eqmod11} impose that the number of items is the maximum capacity of the arcs in the flow. Finally, Constraints \eqref{eqmod16}, \eqref{eqmod17}, \eqref{eqmod18} and \eqref{eqmod19} define the decision variable domains.

\section{Solution methods}
\label{sec:RFO}

As mentioned earlier, to the extent of our knowledge, only \citet{Kang1999} and \citet{Meyr02} approached the ILSSP-NT on parallel machines and introduced heuristic methods. Despite the satisfactory results in both studies, the computational times of the heuristics are compromised by some parameters of the small bucket models, such as the number of micro-periods per macro-period, that may increase the size of the problem and/or require the optimization of several problems with different parameter values.

The best methods related to the ILSSP-NT on parallel machines rely on mathematical programming-based heuristics, such as relax-and-fix (RF) heuristics and fix-and-optimize (FO) heuristics \citep{James2011,Guimaraes2013}. Moreover, intensification approaches such as path-relinking \cite{glover1998tabu} and kernel search \citep{ang07} achieved outstanding results on solving lot sizing problems \citep{NT07,Guastaroba2017,Carvalho2018}. Bearing this in mind, we propose a solution method denoted by RFO, which combines a hybrid of RF and FO. Moreover, we propose two other solution methods referred to as RFO-PR and RFO-KS, which are the RFO combined with intensification heuristics. In both RFO-PR and RFO-KS, RFO searches for initial feasible solutions to the ILSSP-NT on parallel machines. Then, these solutions are used by the path-relinking and kernel search intensification strategies in, respectively, RFO-PR and RFO-KS.

The proposed heuristics, i.e., relax-and-fix (RF), fix-and-optimize (FO), path-relinking (PR) and kernel search (KS) are described at length in the following sections, as well as RFO, RFO-PR and RFO-KS.

\subsection{RF}
\label{sec:RF}

\textcolor{black}{The RF heuristic is a constructive strategy that iteratively solves relaxed mixed integer problems (MIPs) and then fixes the values of some integer variables until the point where all the integer variables are fixed and a feasible solution to the addressed problem is found or an infeasible problem is reached. The first problem solved by a standard RF heuristic is the linear relaxation of the problem. New MIPs are created by iteratively restricting the domain of the decision variables set, which is partitioned into an ordered sequence of $\Phi$ disjoint sets. The initial partition is usually defined according to the solution of the linear relaxation. The order of the sets is basically related to the relevance of their variables, which can be measured by the magnitude of the values in the solution of the linear relaxation of the problem.}

Unlike the classical approach, in lot sizing it is common to design the initial partition of the RF heuristic by considering the sequence of periods, items and/or machines instead of, e.g., the appro\-xi\-mation provided by the linear relaxation solution. Therefore, the variables are assigned to the sets according to the range of the related items, periods and/or machines. For example, let the period be the relevance indicator of the partitioning of the decision variables $x_{iktu}$ into $\Phi$ sets. Each set $S^{\phi}$, $\phi=1,\ldots,\Phi$, is defined as presented in Equation \eqref{Sh}.

\begin{equation}
S^ {\phi} = \{ (i,k,t,u): i \in \{1,n\}, k \in \{1,\ldots,m\}; t \in [t_{\phi},p_{\phi}]; u \in \{t,\ldots,p\}\} \label{Sh}
\end{equation}

\noindent where $t_{\phi}$ and $p_{\phi} \in \{1, \ldots, p\}$.

Given the disjoint sets of variables defined according to some criteria, to solve a certain MIP, the RF heuristic solves one MIP of a sequence of problems at each iteration $v$. In the $v$-th iteration of the RF heuristic, a subproblem of the linear relaxation of the original MIP is solved, where the domain of the variables associated to the tuples $(i,k,t,u)$ from $S^{v}$ is integer, and the domain of the variables associated to the tuples $(i,k,t,u)$ from $S^{v+1}, \ldots,S^{\Phi}$ is real. If $v>1$, the variables associated to the tuples $(i,k,t,u)$ from $S^{1},\ldots,S^{v-1}$ are fixed at integer values from the solutions of MIPs solved at previous iterations. At the first iteration, no variable is fixed in the MIP to be solved. The method halts when a feasible solution to the original problem is found, when it reaches the maximum number of iterations, $\Phi$, or when an MIP is infeasible.

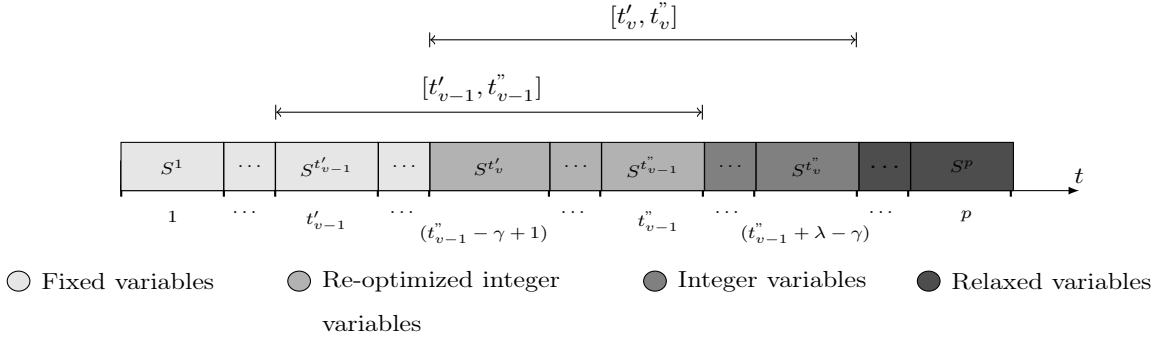
\begin{figure}[!h]
\pgfplotsset{width=10cm,height=2cm,,compat=1.9}
\centering
\begin{tikzpicture}[xscale=0.9,yscale=0.8,transform shape]

\draw [-latex](0,0) coordinate(dd)-- (0,0) coordinate (O1) -- (14,0)coordinate(ff) node[above]{$t$};
\draw [thick] (O1) -- (0,0.5) coordinate(1) -- ++(0,0)coordinate(ff2);

%\draw (-0.5,0.5) node[]{1};\fill[orange] (0,0) circle (1ex)
\draw[fill=black!10!white] (-1.5,-1.5) coordinate(l1) circle (1ex) node[above right=0.0cm and 0.2cm of l1,right,fill=white]{Fixed variables};
\draw[fill=black!30!white] (2.6,-1.5) coordinate(l7) circle (1ex) node[above right=0.0cm and 0.2cm of l7,right,fill=white]{Re-optimized integer};
\draw (2.6,-2.2) coordinate(l8)  node[above right=0.0cm and 0.2cm of l8,right,fill=white]{variables};
\draw[fill=black!50!white] (7.8,-1.5) coordinate(l3) circle (1ex) node[,above right=0.0cm and 0.2cm of l3,right,fill=white]{Integer variables};
\draw[fill=black!70!white] (11.8,-1.5) coordinate(l5) circle (1ex) node[above right=0.0cm and 0.2cm of l5,right,fill=white]{Relaxed variables};

\draw[dashed] (1.5,0) -- (1.5,0|- ff2);
\draw[dashed] (2.25,0) -- (2.25,0|- ff2);
\draw[dashed] (3.75,0) -- (3.75,0|- ff2);
\draw[dashed] (4.5,0) -- (4.5,0|- ff2);
\draw[dashed] (6.25,0) -- (6.25,0|- ff2);
\draw[dashed] (7.0,0) -- (7.0,0|- ff2);
\draw[dashed] (8.5,0) -- (8.5,0|- ff2);
\draw[dashed] (9.25,0) -- (9.25,0|- ff2);
\draw[dashed] (10.75,0) -- (10.75,0|- ff2);
\draw[dashed] (11.5,0) -- (11.5,0|- ff2);
\draw[dashed] (13.0,0) -- (13.0,0|- ff2);

\draw[thick] (0,0.1) -- (0,-0.1) node[below]{};
\draw[thick] (1.5,0.1) -- (1.5,-0.1) node[below]{};
\draw[thick] (2.25,0.1) -- (2.25,-0.1) node[below]{};
\draw[thick] (3.75,0.1) -- (3.75,-0.1) node[below]{};
\draw[thick] (4.5,0.1) -- (4.5,-0.1) node[below]{};
\draw[thick] (6.25,0.1) -- (6.25,-0.1) node[below]{};
\draw[thick] (7.0,0.1) -- (7.0,-0.1) node[below]{};
\draw[thick] (8.5,0.1) -- (8.5,-0.1) node[below]{};
\draw[thick] (9.25,0.1) -- (9.25,-0.1) node[below]{};
\draw[thick] (10.75,0.1) -- (10.75,-0.1) node[below]{};
\draw[thick] (11.5,0.1) -- (11.5,-0.1) node[below]{};
\draw[thick] (13.0,0.1) -- (13.0,-0.1) node[below]{};

\draw (0.75,-0.2) node[below]{\scriptsize{1}};
\draw (1.8,-0.2) node[below]{\scriptsize{$\ldots$}};
\draw (3,-0.2) node[below]{\scriptsize{${t'_{v-1}}$}};
\draw (4.1,-0.2) node[below]{\scriptsize{$\ldots$}};
\draw (5.3,-0.4) node[below]{\scriptsize{$(t^{"}_{v-1}-\gamma +1)$}};
\draw (6.6,-0.2) node[below]{\scriptsize{$\ldots$}};
\draw (7.85,-0.2) node[below]{\scriptsize{$t^{"}_{v-1}$}};
\draw (8.85,-0.2) node[below]{\scriptsize{$\ldots$}};
\draw (10,-0.4) node[below]{\scriptsize{$(t^{"}_{v-1}+\lambda -\gamma)$}};
\draw (11.1,-0.2) node[below]{\scriptsize{$\ldots$}};
\draw (12.35,-0.2) node[below]{\scriptsize{$p$}};

\begin{scope}[shift={(1)}]
\node[ above right=-0.1cm and -0.01cm ,right,draw, minimum width=1.5cm,minimum height=0.8cm,fill=black!10!white](S1){\scriptsize{$S^{1}$}};% {\scriptsize{$S^{1}$}};
%\draw[|<->|,blue] (0,0.8) coordinate(bb1) -- ++(1.5,0) coordinate(bb2) node[above right=0.4cm and 0.4cm of bb1,right,fill=white]{};

\node[right=0.1cm and -0.01cm of S1,right,draw, minimum width=0.75cm,minimum height=0.8cm,fill=black!10!white](S2) {\scriptsize{$\ldots$}};
%\draw[|<->|,blue] (1.5,0.8) coordinate(bb3) -- ++(1.5,0) coordinate(bb4) node[above right=0.4cm and 0.4cm of bb3,right,fill=white]{};%{$S^{t'_{v}}$};

\node[right=0.1cm and -0.01cm of S2,right,draw, minimum width=1.5cm,minimum height=0.8cm,fill=black!10!white](Svi) {\scriptsize{$S^{t'_{v-1}}$}};%{$\ldots$};
%\draw[|<->|,blue] (3,0.8) coordinate(dd1) -- ++(1.5,0) coordinate(dd2) node[above right=0.3cm and 0.4cm of dd1,right,fill=white]{};

\node[right= 0.1cm and -0.01cm of Svi,right,draw, minimum width=0.75cm,minimum height=0.8cm,fill=black!10!white](Sdot1){\scriptsize{$\ldots$}};% {\scriptsize{$S^{t'}$}};
%\draw[|<->|,red] (4.5,0.8) coordinate(bb5) -- ++(1.5,0) coordinate(bb6) node[above right=0.4cm and 0.4cm of bb5,right,fill=white]{};

\node[right= 0.1cm and -0.01cm of Sdot1,right,draw, minimum width=1.75cm,minimum height=0.8cm,fill=black!30!white](Svii){\scriptsize{$S^{t'_{v}}$}};% {\scriptsize{$S^{t'}$}};
%\draw[|<->|,red] (4.5,0.8) coordinate(bb5) -- ++(1.5,0) coordinate(bb6) node[above right=0.4cm and 0.4cm of bb5,right,fill=white]{};

\node[right= 0.1cm and -0.01cm of Svii,right,draw, minimum width=0.75cm,minimum height=0.8cm,fill=black!30!white](Sd){\scriptsize{$\ldots$}};% {\scriptsize{$S^{t'}$}};
%\draw[|<->|,red] (4.5,0.8) coordinate(bb5) -- ++(1.5,0) coordinate(bb6) node[above right=0.4cm and 0.4cm of bb5,right,fill=white]{};

\node[right= 0.1cm and -0.01cm of Sd,right,draw, minimum width=1.5cm,minimum height=0.8cm,fill=black!30!white](Sdot2){\scriptsize{$S^{t^{"}_{v-1}}$}};% {\scriptsize{$S^{t'}$}};
%\draw[|<->|,red] (4.5,0.8) coordinate(bb5) -- ++(1.5,0) coordinate(bb6) node[above right=0.4cm and 0.4cm of bb5,right,fill=white]{};

\node[right=0.1cm and -0.01cm of Sdot2,right,draw, minimum width=0.75cm,minimum height=0.8cm,fill=black!50!white](Sdot3){\scriptsize{$\ldots$}};% {$\ldots$};

\node[right= 0.1cm and -0.01cm of Sdot3,right,draw, minimum width=1.5cm,minimum height=0.8cm,fill=black!50!white](Svf){\scriptsize{$S^{t^{"}_{v}}$}};% {\scriptsize{$S^{(t'+\lambda-1)}$}};
%\draw[|<->|,red] (7.5,0.8) coordinate(bb7) -- ++(1.5,0) coordinate(bb8) node[above right=0.4cm and 0cm of bb7,right,fill=white]{};
\draw[|<->|,black] (2.25,0.8) coordinate(bb7) -- ++(6.25,0) coordinate(bb8) node[above right=0.4cm and 2.0cm of bb7,right,fill=white]{$[t'_{v-1},t^{"}_{v-1}]$};
\draw[|<->|,black] (4.5,2.0) coordinate(bb7) -- ++(6.25,0) coordinate(bb8) node[above right=0.4cm and 2.5cm of bb7,right,fill=white]{$[t'_{v},t^{"}_{v}]$};

\node[right=0.1cm and -0.01cm of Svf,right,draw, minimum width=0.75cm,minimum height=0.8cm,fill=black!70!white](Sdot4){$\ldots$};
%\draw[|<->|,orange] (9,0.8) coordinate(dd3) -- ++(1.5,0) coordinate(dd4) node[above right=0.3cm and 0.5cm of dd3,right,fill=white]{};

\node[right=0.1cm and -0.01cm of Sdot4,right,draw, minimum width=1.5cm,minimum height=0.8cm,fill=black!70!white](Sdot5){\scriptsize{$S^{p}$}};
%\draw[|<->|,orange] (9,0.8) coordinate(dd3) -- ++(1.5,0) coordinate(dd4) node[above right=0.3cm and 0.5cm of dd3,right,fill=white]{};

\end{scope}

\end{tikzpicture}
\caption{\small{Representation of the variable domains associated to the period intervals $[t'_{v-1},t^{"}_{v-1}]$ and $[t'_{v},t^{"}_{v}]$ at iteration $v$.}}
\label{jm:1}
\end{figure}

In \citep{James2011}, time-based partitioning is considered the most appropriate to approach the ILSSP due to the structure of the original MIP the authors use in the RF heuristic. Moreover, different from the aforementioned approach, part of the variables optimized\footnote{ Variables whose domain is integer in a given MIP.} in the MIP to be solved at an iteration $v$ is re-optimized at iteration $v+1$, which means that the variable domains remain integer in the corresponding MIP. Let $\lambda$ be the number of periods whose variable domains must be integer in the MIP of a given iteration $v$. A sequence of period intervals $[t'_{v},t^{"}_{v}]$ is created in such a way that successive intervals share $\gamma < \lambda$ periods, as shown in Figure \ref{jm:1}. In other words, the variables associated to $\gamma$ periods will be re-optimized, i.e., the domain of such variables will also be integer in the MIP of the next iteration.

The strategy suggested in \citep{James2011} is motivated by the use of setup carry-over. The authors claim that disregard the overlapping of intervals when defining the MIPs of the RF heuristic could affect the production sequence once the first setup state of every machine in a given iteration $v$ is fixed at the previous iteration. 

The applied RF is a similar approach to the one presented in \citep{James2011} and relies on overlapping period intervals to define the variable domains along the iterations.

\subsection{FO}
\label{sec:FO}

A classical FO heuristic consists in the successive re-op\-ti\-mi\-za\-tion of simpler MIPs than the original problem, where part of the decision variables is fixed and the other is re-optimized. Different from RF, the FO therefore needs an initial feasible solution whose values will define the values of the variables to be fixed. Similar to the RF heuristic, in the classical FO the decision variables to be fixed are partitioned into $\Phi$ disjoint sets $S^{\phi}$, $\phi = 1, \ldots, \Phi$, which are sorted in order of importance. The importance of the sets is directly related to the relevance of the variables that belong to them. In the FO heuristic, once again, the magnitude of the values in a feasible solution used as input to the method can be used as a criterion to measure the relevance of the variables.

Once the partitioning is defined, the FO heuristic solves an MIP at each iteration $v$, where the variables associated to the tuples from $S^{v}$ are restricted to integer and the remainder variables are fixed at the values obtained for the best solution found up until that moment. The method halts after solving a sequence of $\Phi$ MIPs.

This strategy has been extensively explored to solve MIPs found in the literature  \citep{Pochet2006,Sahling2009,Helber2010,James2011,Guimaraes2013}, in particular, to approach lot sizing problems with excellent results \cite{James2011,Guimaraes2013}. Bearing this in mind and based on the aforementioned methodology, we apply an FO heuristic to improve the solution found by the RF heuristic. \textcolor{red}{Next, we present the steps of the proposed RF and FO and, then, the proposed RFO heuristic.}

\subsection{RFO heuristic}

\textcolor{red}{ RF and FO have virtually the same set of steps. For this reason, we present the methods together, for a better reading. It is important to highlight that the RF and FO steps are not performed simultaneously. Moreover, two of the six steps only occur in the FO heuristic,  informed next and marked with~`(*)'.}

\begin{enumerate}
 \item \textit{Definition of parameters and time-based variable partitioning} \\
	In both  RF and FO heuristics, the set of decision variables $y$ is partitioned into $p$ disjoint sets $\{y_{ijkt}: (i,j,k,t) \in S^{u}\}, \; u = 1, \ldots, p$, where set $S^{u}$ is composed of every tuple $(i,j,k,u)$. To define the MIP of a given iteration $v$, let $\lambda$  be a constant value that represents the number of periods whose variable domains are integer in the RF \textcolor{red}{and FO heuristics}. Consider also \textcolor{red}{$\gamma<\lambda$ an integer value that describes} the number of periods whose variable domains were integer in previous iterations and remained integer at iteration $v$ rather than being fixed in the MIP of the current iteration. Therefore, the period interval $[t'_{v},t^{"}_{v}]$ indicates the periods under investigation at iteration $v$, i.e., the domain of the variables associated to such periods must be integer.
		
	\textcolor{red}{Let $\theta$ be a constant value, where $\theta = \lceil(p-\lambda) /(\lambda -\gamma)\rceil+1$. On the one hand, $\theta$ indicates the last iteration counter in the RF heuristic. On the other, $\theta$ indicates the last iteration counter of a cycle of iterations in the FO heuristic\footnote{If the FO heuristic finds a new solution that is strictly better than the current best one after completing a cycle of iterations, we attempt to improve such a solution, by considering a new cycle of the FO heuristic having it as starting solution}. The methods employ $\theta$ to define the intervals $[t'_{v},t^{"}_{v}]$, as follows:  (i)  $t'_{1}=1$ and $t^{"}_{1} = t'_{1}+ \lambda -1$; (ii) for $v \in \{2,3,\ldots, \theta-1\}$, $t'_{v}=t^{"}_{v-1}- \gamma + 1$ and $t^{"}_{v} = t'_{v} + \lambda -1$; and (iii)   $t'_{\theta}=p- \gamma + 1$ and $t^{"}_{\theta} = p$.}

    Let \textsf{TimeLimit-RF} and \textsf{TimeLimit-FO} be the maximum available time to perform the RF and FO heuristics, respectively. To ensure that the methods do not stop before the end of the last iteration of a cycle of the methods, we impose the respective time limits to solve each MIP: i) \textsf{MIP-RF}= \textsf{TimeLimit-RF}/ $\theta$, for RF; and ii) \textsf{MIP-FO}= \textsf{TimeLimit-FO}/$\theta$, for FO. For the FO heuristic, a parameter called \textsf{ElapsedTime-FO}, initialized with the value 0, keeps track of the time spent up to the end of a given iteration to update \textsf{MIP-FO} in the next iteration.

    \item \textit{Solve the \textcolor{red}{MIP of iteration $v$}}  
    
    \textcolor{red}{The following MIP must be solved at iteration $v$ of the RF and FO heuristics, \textcolor{red}{within the given time limits, \textsf{MIP-RF} and \textsf{MIP-FO}, respectively}. The MIPs differ by Heuristic Constraints \eqref{rfeq:heur}, which are described next for each method.}

        \begin{align}
        (MIP)^{v}:  & &\quad   \min \quad  \mathop {\sum }\limits_{i=1}^{n} \mathop {\sum }\limits_{k=1}^{m} \mathop {\sum }\limits_{t=1}^{p}\mathop {\sum }\limits_{u=t}^{p}(u-t)h_{i}x_{iktu} + \mathop {\sum }\limits_{i=1}^{n} \mathop {\sum }
        \limits_{j=1}^{m}\mathop {\sum }\limits_{k=1}^{m}\mathop {\sum }\limits_{t=1}^{p}c_{ijk}y_{ijkt} & \label{rf:1}
        \end{align}

        subject to:
        \begin{align}
        & \mbox{Constraints } \eqref{eqmod1} - \eqref{eqmod17}, & & \label{rfeq:1} \\
        & \mbox{Heuristic Constraints} & & \label{rfeq:heur}
        \end{align}

    \textcolor{red}{Let $y^{(MIP)^{v-1}}_{ijkt}$ be the value associated to variable $y_{ijkt}$ in the solution of problem $(MIP)^{v-1}$ solved in the previous iteration of the method (either RF or FO).}  
    \textcolor{red}{In the RF heuristic, the Heuristic Constraints \eqref{rfeq:heur} are:}
    
        \begin{align}
    	&z_{ikt} \in [0,1], x_{iktu} \geq 0, & \forall & \; (i,k,t,u) & \label{eqM_1} \\
    	&y_{ijkt} = y^{(MIP)^{v-1}}_{ijkt} ,  & \forall & \; y_{ijkt}: (i,j,k,t) \in  \mathop{\bigcup}\limits_{\phi =1}^{t'_{v}-1} S^{\phi} & \label{eqM_2}\\
    	&y_{ijkt} \in \{0,q_{ikt}\},   & \forall & \; y_{ijkt}: (i,j,k,t) \in \mathop{\bigcup}\limits_{\phi =t'_{v}}^{t^{"}_{v}} S^{\phi} & \label{eqM_3}\\
    	&y_{ijkt} \in [0,q_{ikt}],   & \forall & \; y_{ijkt}: (i,j,k,t) \in  \mathop{\bigcup}\limits_{\phi =t^{"}_{v}+1}^{p} S^{\phi} & \label{eqM_4}
        \end{align}
        
	\textcolor{red}{Note that} at the first iteration, Constraints \eqref{eqM_2} are not considered in problem $(MIP)^{1}$ since there is no variable to be fixed.
	
	\textcolor{red}{To define the Heuristic Constraints \eqref{rfeq:heur} in the FO heuristic,} let $(x^{*},y^{*},z^{*})$ and $(S^ {1}, \ldots, S^ {p})$ be the best feasible solution found up to the moment and the partitioning defined in Step 1, respectively. \textcolor{red}{Therefore, in the FO heuristic, the Heuristic Constraints \eqref{rfeq:heur} are:}
		
	\begin{align}
	    & \mbox{Constraints }  \eqref{eqmod18}, & & & \label{fo:1}\\
		&y_{ijkt} = y_{ijkt}^{*} ,  & \forall & \; y_{ijkt}: (i,j,k,t) \in \mathop{\bigcup}\limits_{\phi =t'_{v}}^{t^{"}_{v}} S^{\phi} & \label{eqM2_2}\\
		&y_{ijkt} \in \{0,q_{ikt}\}, & \forall & \; y_{ijkt}: (i,j,k,t) \in \mathop{\bigcup}\limits_{\phi =1, \phi \neq \{t'_{v},\ldots,t^{"}_{v}\}}^{p} S^{\phi} & \label{eqM2_3}
	\end{align}

	\item \textit{Update the best solution$^{(*)}$} \\
     \textcolor{red}{This updating is exclusive of the FO heuristic.} In this case, let $Z^{*}$ and $Z^{(MIP)^{v}}$ be the objective function values of the overall best solution and of the solution to $(MIP)^{v}$, respectively. Solution $(x^{*},y^{*},z^{*})$ is updated with solution $(x^{(MIP)^{v}},y^{(MIP)^{v}},z^{(MIP)^{v}})$, in case $Z^{*} > Z^{(MIP)^{v}}$.

    \item \textit{Update $v$} \\
		\textcolor{red}{In the RF heuristic,} after solving $(MIP)^{v}$, the value of $v$ is incremented by 1. 
        \textcolor{red}{In the FO  heuristic, if} $v<\theta$, the value of $v$ is incremented by 1. Otherwise, if an improved solution was found at any iteration of the current cycle, when $v$ reaches the value $\theta$, $v$ is updated with value 1. If no improved solution was found by the end of a cycle, when $v = \theta$, $v$ is fixed at value 0.
      
	\item \textit{Update \textup{\textsf{ElapsedTime-FO} and \textup{\textsf{MIP-FO}}$^{(*)}$}}

	\textsf{ElapsedTime-FO} is updated with the computational time spent up to this point of the FO heuristic. Then, \textsf{MIP-FO} is updated by making \textsf{MIP-FO}$\, = \,($\textsf{TimeLimit-FO} $-$ \textsf{ElapsedTime-FO}$)/(\theta-v+1)$, which is the available time divided by the number of remaining iterations to complete a cycle.
    
    \item \textit{Stop criteria} \\
         RF stops if $v = \theta$ or if problem $(M1)^{v}$ is infeasible. Otherwise, the heuristic repeats Steps 2 to 4.
        If either \textsf{ElapsedTime-FO} is greater than or equal to \textsf{TimeLimit-FO} or no improved solution is found by the end of a cycle, the algorithm stops and returns $(x^*,y^*,z^*)$. Otherwise, the heuristic repeats Steps 2 to 6.

\end{enumerate}

\textcolor{red}{In the proposed RFO, i}f the RF heuristic returns a feasible solution to the ILSSP-NT on parallel machines, we apply the FO heuristic to the solution. \textcolor{red}{Appendix A shows pseudo-codes of the RF and FO heuristics. Algorithm~\ref{alg:RFO} shows a pseudo-code of the proposed RFO.} 

\begin{algorithm}[!h]
	\begin{small}
 		%\SetAlgoNoEnd
		\SetAlgoVlined
 		%\SetAlgoLined
 		\KwData{Instance, \textsf{TimeLimit-RF},  \textsf{TimeLimit-FO}.}
 		\KwResult{\textcolor{black}{Solutions $(x^{RF},y^{RF},z^{RF})$, $(x^{RF^*},y^{RF^*},z^{RF^*})$ and $(x^{FO},y^{FO},z^{FO})$} or warning message.}    %($x^{RF},y^{RF},z^{RF}$)
		\textcolor{black}{$\{(x^{RF},y^{RF},z^{RF})$, $(x^{RF^*},y^{RF^*},z^{RF^*})\}$ $\leftarrow$ RF heuristic(\textsf{TimeLimit-RF});} \\
		\textcolor{black}{\textsf{TimeLimit-FO} $\leftarrow$ \textsf{TimeLimit-FO} $+$ \textsf{AvailTime-RF};} \\
		\If{ \textcolor{black}{$(x^{RF^*},y^{RF^*},z^{RF^*})$} is feasible}{
		    \textcolor{black}{$(x^{FO},y^{FO},z^{FO})$} $\leftarrow$ FO heuristic(\textcolor{black}{$(x^{RF^*},y^{RF^*},z^{RF^*})$}, \textsf{TimeLimit-FO}); \\
		    Return \textcolor{black}{$\{(x^{RF},y^{RF},z^{RF})$, $(x^{RF^*},y^{RF^*},z^{RF^*}), (x^{FO},y^{FO},z^{FO})\}$}.
		}{
            Return warning message: ``The method was not able to find a feasible solution to the ILSSP-NT on parallel machines''.
		}
 		\caption{RFO heuristic}\label{alg:RFO}
 	\end{small}
\end{algorithm}

Algorithm \ref{alg:RFO} presents a hybrid of the RF and FO heuristics,  named RFO, to solve the ILSSP-NT. \textcolor{black}{It starts by calling the RF heuristic that returns \textcolor{red}{a pair of solutions. The first solution, represented by $(x^{RF},y^{RF},z^{RF})$, is} the solution obtained in \textcolor{red}{iteration $\theta-3$ of the RF heuristic, if the method reaches such an iteration. If a feasible solution is found in an earlier iteration, it is assigned to $(x^{RF},y^{RF},z^{RF})$. The reason to keep $(x^{RF},y^{RF},z^{RF})$ is better explained in the next section, in the first step of the PR description. The second solution, called $(x^{RF^*},y^{RF^*},z^{RF^*})$,  is  the best solution found by the RF heuristic. } Then,  parameter  \textsf{TimeLimit-FO} is updated if the RF heuristic spent less than \textsf{TimeLimit-RF}. In this case, we add to \textsf{TimeLimit-FO} the spare time of the RF heuristic, called  \textsf{AvailTime-RF}. If $(x^{RF^*},y^{RF^*},z^{RF^*})$ is feasible, the FO heuristic uses it as starting solution to obtain an improved solution $(x^{FO},y^{FO},z^{FO})$, which is returned by the method along with $(x^{RF},y^{RF},z^{RF})$  and $(x^{RF^*},y^{RF^*},z^{RF^*})$. Otherwise, the RFO heuristic returns a warning message indicating that it was not able to find a feasible solution for the problem.}

To improve the best solution found by RFO, we propose an improvement step through a PR and a KS heuristic. The PR strategy is performed only if the RFO heuristic provides two  feasible solutions with distinct objective function values. To perform the KS heuristic, at least one feasible solution must be found by the RFO heuristic. The PR and KS heuristics are described at length in the following sections.

%%%%%%%%%%%%%%%%%%%%%%%%%%%%%%%%%%%%%%%%%%%%%%%%%%%%%%%%%%%%%%%%%%%%%%%%%%%

\subsection{PR heuristic}
\label{sec:PR}

Path-relinking (PR) \citep{glover1998tabu} is an intensification strategy that uses a feasible solution as the starting point of a sequence of solutions, which progressively gets closer to another solution defined as guiding solution.  PR is usually proposed conjointly with a diversification-based heuristic that provides the set of solutions in-between the sequence of solutions, also known as paths.

In the standard form of the PR heuristic, the strategy uses an elite set of solutions, from which the \emph{initial} and \emph{guiding} (final) solutions are selected. Then, to perform a path between these two solutions, it is necessary to gradually add characteristics of the guiding solution to a copy of the initial solution. Variants of PR exploit different approaches to build such paths in \citep{GL00}.

\citet{James2011} demonstrated that the performance of a method to solve ILSSP-NT is strongly related to the local search embedded in the strategy. Moreover, the PR heuristics proposed in \citep{NT07, Carvalho2016} obtained outstanding results when solving lot sizing problems. Together, these two facts make up the reasons why we employ in our solution method such a heuristic\textcolor{black}{, described next}.
 
\begin{enumerate}
\item \textit{Define parameters and elite set}

Let $\mathcal{E}$ be the elite set of solutions composed of the two solutions found by the RF heuristic and one solution obtained by the FO heuristic. \textcolor{red}{The first solution, referred to as ($x^{\mathcal{E}(1)},y^{\mathcal{E}(1)},z^{\mathcal{E}(1)}$), is the solution obtained by the RF heuristic in the iteration where the variables associated to the periods $t'_{\theta-2}$ to $p$ have their domain relaxed. Therefore, there is no guarantee that this solution is feasible. The second solution of $\mathcal{E}$, referred to as ($x^{\mathcal{E}(2)}, y^{\mathcal{E}(2)}, z^{\mathcal{E}(2)}$), is the best solution between those found by the RF and PR heuristics. It is worth mentioning that if ($x^{\mathcal{E}(1)},y^{\mathcal{E}(1)},z^{\mathcal{E}(1)}$) is feasible, it will coincide with ($x^{\mathcal{E}(2)}, y^{\mathcal{E}(2)}, z^{\mathcal{E}(2)}$). The third solution, referred to as ($x^{\mathcal{E}(3)}, y^{\mathcal{E}(3)}, z^{\mathcal{E}(3)}$), is the best solution obtained by the FO heuristic.}

\textcolor{red}{Along the iterations, the second solution is replaced by a better quality solution, if it is found during the search process. The objective function value of an infeasible solution is calculated using Equation~\eqref{rf:1}, with no penalty in the violated constraints. The initial solution in the proposed PR is ($x^{\mathcal{E}(3)}, y^{\mathcal{E}(3)}, z^{\mathcal{E}(3)}$), whereas the other two solutions of $\mathcal{E}$ are used as guiding solutions. On the one hand, by defining ($x^{\mathcal{E}(1)},y^{\mathcal{E}(1)},z^{\mathcal{E}(1)}$) as a guiding solution, we aim at allowing a better diversification of the method. On the other, our goal in employing ($x^{\mathcal{E}(2)}, y^{\mathcal{E}(2)}, z^{\mathcal{E}(2)}$) and ($x^{\mathcal{E}(3)}, y^{\mathcal{E}(3)}, z^{\mathcal{E}(3)}$) in PR is to ensure the intensification in the neighborhood of the feasible solutions found by the RF and FO heuristics.}

\textcolor{red}{To restrict the periods whose associated variables are candidates to be fixed at 0, we define a parameter \textcolor{red}{$\zeta\leq p$. This means that we will never fix the variables associated with the last $ p-\lceil\zeta\rceil$ periods of the problems. Its tuning, also considering $\zeta=p$, which corresponds to the classical methodology, is discussed in Section~\ref{sec:tuning}.} }

One of the stop criteria of the PR heuristic is the time limit, given by parameter \textsf{TimeLimit-PR}. \textsf{MIP-PR} is another time-related parameter, set with half of the value of \textsf{TimeLimit-PR}, used to define the maximum time to solve the problem described in Step \textcolor{red}{3}. The time spent up to a point of the PR heuristic is denoted by \textsf{ElapsedLimit-PR} and starts with the value 0.

\item \textit{\textcolor{red}{Define set with shortcut items}}

\textcolor{red}{The introduced PR strategy is specially tailored to approach the ILSSP-NT. For this reason, we  define a set $\Psi$, $\Psi \subset \{1, \ldots, n\}$, that contains all items that can be used to diminish the costs and times of cleansing in setup changeovers, i.e., the shortcut items.} 

\item \textit{Solve \textcolor{red}{$(M1)$}}

Given the solutions in $\mathcal{E}$, the PR strategy attempts to find a new feasible solution by solving problem $M1$, i.e., formulation \eqref{nt:fo}-\eqref{eqmod19} also subject to Constraints \eqref{preq:3}.

    \begin{align}
	& y_{ijkt} = 0 , \mbox{ if } (y_{ijkt}^{\mathcal{E}(1)} \wedge y_{ijkt}^{\mathcal{E}(2)} \wedge y_{ijkt}^{\mathcal{E}(3)}) \leq 0.001, & \forall & \; (i,j,k,t), \, \{i, j\} \notin \Psi, \, \textcolor{red}{t < \zeta} & \label{preq:3} 
	\end{align}

Constraints \eqref{preq:3} assign the null value to all variables $y_{ijkt}$, such that $i,j\notin \Psi$,  \textcolor{red}{$t < \zeta$}, and whose $y_{ijkt}^{\mathcal{E}(1)}$, $y_{ijkt}^{\mathcal{E}(2)}$ and $y_{ijkt}^{\mathcal{E}(3)}$ are at most 0.001. 

To solve $(M1)$ we used the CPLEX solver time limited to \textsf{MIP-PR}. The solution to $(M1)$ is referred to as ($x^{(M1)}, y^{(M1)}, z^{(M1)}$).

\item \textit{Update elite set}

Let $Z^{\mathcal{E}(2)}$ and $Z^{(M1)}$ be the objective function values of ($x^{\mathcal{E}(2)},y^{\mathcal{E}(2)},z^{\mathcal{E}(2)}$) and \linebreak ($x^{(M1)}, y^{(M1)}, z^{(M1)}$), respectively. Solution ($x^{\mathcal{E}(2)}, y^{\mathcal{E}(2)}, z^{\mathcal{E}(2)}$) is updated with solution ($x^{(M1)},\linebreak y^{(M1)},z^{(M1)}$), in case $Z^{\mathcal{E}(2)} > Z^{(M1)}$. 

\item \textit{Update parameters \textup{\textsf{ElapsedTime-PR}} and \textup{\textsf{MIP-PR}}}

After solving $(M1)$, \textsf{ElapsedTime-PR} is updated with the time consumed by the PR heuristic up to this point. Then, \textsf{MIP-PR} is set to (\textsf{TimeLimit-PR} $-$ \textsf{ElapsedTime})$/2$, which is half of the time available so far.

\item \textit{Stop criteria}

If either \textsf{ElapsedTime-PR} is greater than or equal to \textsf{TimeLimit-PR} or the objective function values $Z^{\mathcal{E}(2)}$ and $Z^{\mathcal{E}(3)}$ respectively associated to the solutions ($x^{\mathcal{E}(2)}, y^{\mathcal{E}(2)}, z^{\mathcal{E}(2)}$) and ($x^{\mathcal{E}(3)}, y^{\mathcal{E}(3)}, z^{\mathcal{E}(3)}$) are equal, the algorithm stops and returns $\mathcal{E}$. Otherwise,  the heuristic repeats Steps 3 to 6.

\end{enumerate}

Algorithm \ref{alg:PR} presents a pseudo-code for the proposed PR heuristic, where the elite set with the three solutions is given as input.

\begin{algorithm}[!h]
	\begin{small}
 		%\SetAlgoNoEnd
		\SetAlgoVlined
 		%\SetAlgoLined
 		\KwData{$\mathcal{E}$, \textsf{TimeLimit-PR}.}
 		\KwResult{Updated solution set $\mathcal{E}$.}    %($x^{RF},y^{RF},z^{RF}$)
 		Define set $\Psi$; \\
 		\textsf{MIP-PR} $\leftarrow$ \textsf{TimeLimit-PR}$/2$; \\
        \textsf{ElapsedTime-PR} $\leftarrow$ 0; \\
 		\While{\textup{\textsf{ElapsedTime-PR} $< $ \textup{\textsf{TimeLimit-PR}} and $Z^{\mathcal{E}(2)} \neq Z^{\mathcal{E}(3)}$}}{
        	$(x^{(M1)},y^{(M1)},z^{(M1)})$ $ \leftarrow $ solution of the problem $(M1)$ within the time limit of \textsf{MIP-PR};\\
	        \If{$Z^{(M1)} < Z^{\mathcal{E}(2)}$}{
	            $(x^{\mathcal{E}(2)},y^{\mathcal{E}(2)},z^{\mathcal{E}(2)}) \leftarrow (x^{(M1)},y^{(M1)},z^{(M1)})$; \\
	        }
        	Update \textsf{ElapsedTime-PR}; \\
            \textsf{MIP-PR} $\leftarrow($\textsf{TimeLimit-PR} - \textsf{ElapsedTime-PR}$)/2$. \\
        }
        Return $\mathcal{E}$. \\
 		\caption{PR heuristic}\label{alg:PR}
 	\end{small}
\end{algorithm}

To solve the ILSSP-NT, we propose a hybrid solution method, named RFO-PR, that combines the RF, FO and the PR heuristics. We present a pseudo-code of the RFO-PR heuristic in Algorithm \ref{alg:RFO-PR}.

\begin{algorithm}[!h]
	\begin{small}
 		%\SetAlgoNoEnd
		\SetAlgoVlined
 		%\SetAlgoLined
 		\KwData{Instance, \textcolor{black}{$(x^{RF},y^{RF},z^{RF})$, $(x^{RF^*},y^{RF^*},z^{RF^*})$, $(x^{FO},y^{FO},z^{FO})$, \textsf{AvailTime-RFO}, \textsf{TimeLimit-PR}}.}
 		\KwResult{A solution to the problem or a warning message.}    %($x^{RF},y^{RF},z^{RF}$)
		\textcolor{black}{$(x^{\mathcal{E}(1)},y^{\mathcal{E}(1))},z^{\mathcal{E}(1)})$ $\leftarrow$ $(x^{RF},y^{RF},z^{RF})$;} \\
		\textcolor{black}{$(x^{\mathcal{E}(2)},y^{\mathcal{E}(2))},z^{\mathcal{E}(2)})$ $\leftarrow$ $(x^{RF^*},y^{RF^*},z^{RF^*})$;} \\
		\textcolor{black}{$(x^{\mathcal{E}(3)},y^{\mathcal{E}(3))},z^{\mathcal{E}(3)})$ $\leftarrow$ $(x^{FO},y^{FO},z^{FO})$;} \\
		\textcolor{black}{\textsf{AvailTime-PR} $\leftarrow$ \textsf{TimeLimit-PR} + \textsf{AvailTime-RFO};} \\
	    \While{\textcolor{black}{\textup{\textsf{AvailTime-PR}} $> 0$ and $Z^{\mathcal{E}(2)} \neq Z^{\mathcal{E}(3)}$}}{
	    	\textcolor{black}{$Z^{PR} \leftarrow Z^{\mathcal{E}(2)}$;} \\
	    	\textcolor{black}{$\mathcal{E} \leftarrow$ PR heuristic($\mathcal{E}$, \textsf{AvailTime-PR});} \\
	    	\textcolor{black}{Update \textsf{AvailTime-PR};} \\
		    \If{\textcolor{black}{\textup{\textsf{AvailTime-PR}} $> 0$ and $Z^{PR} > Z^{\mathcal{E}(2)}$ and $Z^{\mathcal{E}(2)} \neq Z^{\mathcal{E}(3)}$} }{
		    	\textcolor{black}{$(x^{\mathcal{E}(3)},y^{\mathcal{E}(3)},z^{\mathcal{E}(3)})$ $\leftarrow$ FO heuristic($(x^{\mathcal{E}(2)},y^{\mathcal{E}(2)},z^{\mathcal{E}(2)})$, \textsf{AvailTime-PR});} \\
		    	\textcolor{black}{Update \textsf{AvailTime-PR};} \\
		    }			    
	    }
	    \eIf{\textcolor{black}{$Z^{\mathcal{E}(2)}$ or $Z^{\mathcal{E}(3)}$ is feasible}}
	    {
    	    \eIf{\textcolor{black}{$Z^{\mathcal{E}(2)} < Z^{\mathcal{E}(3)}$}}
    	    {
    		    \textcolor{black}{Return $(x^{\mathcal{E}(2)},y^{\mathcal{E}(2)},z^{\mathcal{E}(2)})$.} \\
    		}{
    	    	\textcolor{black}{Return $(x^{\mathcal{E}(3)},y^{\mathcal{E}(3)},z^{\mathcal{E}(3)})$.} \\
    	    }
	    }{
            Return warning message indicating that the method has not found a feasible solution to the ILSSP-NT on parallel machines.
		}
 		\caption{RFO-PR heuristic}\label{alg:RFO-PR}
 	\end{small}
\end{algorithm}

Note that, in Algorithm \ref{alg:RFO-PR}, in lines 1 to 3, the elite set to be used by the PR heuristic is built using the solutions of the RFO heuristic given as input. After that, the maximum running time of  RFO-PR, referred to as \textsf{AvailTime-PR}, is set with the imposed time limit  to the PR heuristic -- \textsf{TimeLimit-PR} -- plus \textsf{AvailTime-RFO}, which is the spare time of the RFO heuristic. In the sequence, the main loop of the RFO-PR heuristic starts. In the loop, the PR and FO heuristics are performed respectively while \textsf{AvailTime-PR} is greater than zero and the objective function values associated with  $(x^{\mathcal{E}(2)},y^{\mathcal{E}(2)},z^{\mathcal{E}(2)})$ and $(x^{\mathcal{E}(3)},y^{\mathcal{E}(3)},z^{\mathcal{E}(3)})$ are different. If a solution  with a better objective function value in comparison to $(x^{\mathcal{E}(2)},y^{\mathcal{E}(2)},z^{\mathcal{E}(2)})$ is found by the PR heuristic, the FO heuristic is applied to such a solution. The solution found by the FO heuristic is assigned to $(x^{\mathcal{E}(3)},y^{\mathcal{E}(3)},z^{\mathcal{E}(3)})$ whenever this heuristic is called. Finally, in lines 12 to 18, the RFO-PR heuristic returns the best solution found or a warning message indicating that the method has not found a feasible solution to the problem.

\subsection{KS heuristic}
\label{sec:KS}

The main idea behind KS consists, basically, in reducing an original integer problem by re-optimizing the variables most likely to be in the optimal solution and fixing the remaining variables at 0. This means that the reduced problem does not consider part of the decision variables, i.e., it considers just a set of variables denoted by \emph{kernel}.  

The standard framework of a KS has two phases known as Initialization and Extension phases. In the Initialization phase, the method decides which variables have the best chance of belonging to the optimal solution of the problem and builds the main structure of a KS, which is a kernel and a sequence of buckets. In an attempt to guide the decision on the most promising variables for the optimal solution, KS strategies usually employ a linear relaxation of the problem under study \citep{ang07,angelelli2010}. In line with this, in a minimization problem, the variables with the reduced costs less than a threshold are assigned to the kernel. The remaining variables are arranged as a sequence of equally-sized $nb$ buckets organized in order of reduced costs. Once the kernel and the sequence of buckets is created, a reduced subproblem is solved by considering the re-optimization of the variables in the kernel and fixing the variables assigned to the buckets at 0.

The Extension phase is responsible for investigating the solutions of the kernel problem modified by not considering the values of the variables of each bucket to be fixed. The investigation is then performed iteratively by solving, at each iteration $v$, where $v \in \{1, \ldots, nb\}$, a subproblem restricted to the kernel and the $v$-th bucket and by fixing at 0 all the remaining variables. Moreover the subproblem is also subject to two additional constraints. The first constraint limits the objective function value to an upper bound in order to restrict the solution space and provide a faster search. The second constraint is responsible for the enlargement of the kernel by considering to add at least one promising variable from the bucket investigated at iteration $v$ to the kernel in an attempt to obtain a better solution. If a feasible solution is found at iteration $v$, the variables from the $v$-th bucket whose values in the solution are greater than 0 are added to the kernel. The Extension phase halts when all the $nb$ buckets have been investigated.

In this paper, we propose a KS specially designed to approach the ILSSP-NT, whose definition of the kernel and the buckets depend on the items with cleansing properties. Solutions found by the RF and FO heuristics also guide the construction of the kernel and the buckets, the reason we named as RFO-KS the introduced solution method. Next, we thoroughly describe both the Initialization and the Extension phases of the RFO-KS heuristic.

\subsubsection*{\textbf{Initialization phase}}
\label{sec:PI}

Different from most KSs found in the literature, besides using the reduced costs related to the variables of the linear relaxation of the problem, a feasible solution obtained by the RF and FO heuristics \textcolor{black}{and the cleansing properties of some items} also guide the construction of the kernel and the buckets. The steps of this phase are described next.

\begin{enumerate}
	\item \textit{Define parameters} \\
	Let $(x^{RF^*},y^{RF^*},z^{RF^*})$ and $(x^{FO},y^{FO},z^{FO})$ be the best solutions found by the RF and FO heuristics, respectively. The best solution obtained by the KS heuristic is represented by $(x^{*},y^{*},z^{*})$ and initialized with the solution $(x^{FO},y^{FO},z^{FO})$.
	
	The maximum time for the KS heuristic to search for solutions is given by parameter \textsf{AvailTime-KS} that is the sum of two parameters: a pre-defined value in seconds called \textsf{TimeLimit-KS} and  the spare time of the RFO heuristic. A  parameter called \textsf{MIP-KS}, set with one-third of the value of \textsf{AvailTime-KS}, defines the maximum amount of time to solve each reduced subproblem introduced in the next step of this phase.
	
	\textcolor{red}{The set of shortcut items defined as $\Psi$ in the PR heuristic will be used in the KS strategy to guide the construction of the kernel and the buckets, as discussed in the third step of the method.}

    \item \textit{Solve the linear relaxation of the ILSSP-NT on parallel machines} \\
    \textcolor{red}{The method solves problem \eqref{nt:fo}-\eqref{eqmod19}, denoted here to as $(LP)$, with all integrality constraints relaxed.}    The optimal solution to $(LP)$ and the reduced costs associated with its variables are represented by tuples $(x^{(LP)},y^{(LP)},z^{(LP)})$ and $(rc^{x},rc^{y},rc^{z})$, respectively.
    
    \item \textit{Build the kernel and sequence of buckets}\\
    Let $\Delta = (XZ,K)$ be the set responsible for keeping the variables of the kernel, where $XZ$ comprises variables $x_{iktu}$ and $z_{ikt}$, $\forall \, i,k,t,u$. $K$ is composed of the setup state variables $y_{ijkt}$ $\forall\{i,j\} \in \Psi$ plus the variables that satisfy at least one of the following conditions: (i) associated reduced costs $rc^{y}_{ijkt}$ are lower than or equal to 0; (ii) $y_{ijkt}^{(LP)}$ is greater than 0; (iii)  either $y_{ijkt}^{RF^*}$ or $y_{ijkt}^{FO}$ is equal to 1. The complementary set $\bar{K}$ with all the variables $y_{ijkt}$ that are not in the kernel are sorted in increasing order according to their associated values in the relaxed solution $y^{(LP)}_{ijkt}$. Next, the variables from $\bar{K}$ are evenly distributed into $nb$ buckets, considering the priority of indexes $i\rightarrow j \rightarrow k \rightarrow t$. Therefore, $\lceil|\bar{K}|/nb \rceil$ binary variables are assigned to each bucket $B_{v}$, where $v \in \{1,\ldots,nb\}$.
    
    \item \textit{Solve the MIP of the ILSSP-NT on parallel machines restrict to $\Delta$}\\
    \textcolor{red}{The method searches for the first feasible solution by solving problem $(M2)$, whose formulation is the same as $(LP)$,  subject to two additional constraints: (i) one that ensures that the solution quality of the problem is lower than or equal to the value of the objective function of solution $(x^{FO},y^{FO},z^{FO})$ returned by RFO; and (ii) a set of constraints that fixes at 0  all variables $y_{ijt}$ that belong to the set of unpromising variables $\bar{K}$.}
      
	\item \textit{Update the best solution}\\
        If a feasible solution to problem $(M2)$ is found by CPLEX within the time limit of \textsf{MIP-KS} seconds, the method updates the best solution $(x^{*},y^{*},z^{*})$.
\end{enumerate}

\subsubsection*{\textbf{Extension phase}}
\label{sec:FE}

The Extension phase iteratively exploits a subproblem restrict to set $\Delta$ combined with the $v$-th bucket, where $v$ represents the current iteration. In \citep{Guastaroba2017}, after testing different configurations to find a good value to the number of buckets $nb$, the authors conclude that $nb$ equals to the number of elements in the complementary set $\bar{K}$ divided by the number of elements in the kernel $K$ leads to a good trade-off between the solution quality and the algorithm efficiency. Therefore, in this paper, we use $\lceil|\bar{K}|/|K|\rceil$ to define the number of buckets to be investigated during the Extension phase of the proposed KS heuristic.

\begin{enumerate}
	\item \textit{Update \textup{\textsf{MIP-KS}}}\\
	   At each iteration $v$, starting from 1 and ending at $nb$, the updating of parameter \textsf{AvailTime-KS} occurs. This updating is required to calculate the \textsf{MIP-KS} value, which is the division of \textsf{AvailTime-KS} by the number of remaining iterations of the Extension phase. This ensures that the method goes over all buckets.
	
	\item \textit{Solve subproblem restrict to $K\cup B_v$}\\
        \textcolor{red}{At each iteration $v$, the method solves subproblem $(M3)^{v}$, whose formulation is the same as $(LP)$,  subject to three additional constraints: (i) one that ensures that the objective function is lower than or equal to the best solution found so far; (ii) a set of constraints that fixes at 0 all variables $y_{ijt}$ that belong to the set of unpromising variables $\left(\mathop \bigcup \limits_{h=1,h \neq v}^{nb}B_h\right)$; and (iii) a constraint which demands that at least one of the variables from bucket $B_v$ must belong to the solution of subproblem $(M3)^{v}$.}
	
	\item \textit{Update the best solution and kernel} \\
        If a feasible solution to the subproblem $(M3)^{v}$ is found by CPLEX within the time limit of \textsf{MIP-KS} seconds, the method updates the solution $(x^{*},y^{*},z^{*})$ and the kernel.
        
        To update the kernel let $B_{v}^+$ be the set with the promising setup state variables with value 1 in the solution of $(M3)^{v}$. Then, the enlargement criterion of the kernel is performed by adding $B_{v}^+$ into $K$, i.e., $K := (K \cup B_{v}^+)$. Moreover, $B_{v}$ is updated by making $B_{v} := B_{v} \setminus B_{v}^{+}$.
	
	\item \textit{Stop criterion} \\
    	The value of $v$ is incremented by 1. If $v \leq nb$, the Extension phase repeats Steps 1 to 4. Otherwise, the Extension phase halts and the RFO-KS heuristic returns solution $(x^{*},y^{*},z^{*})$ \textcolor{black}{ or a message warning that no feasible solution has been found}.
\end{enumerate}

A pseudo-code of RFO-KS is presented in Algorithm \ref{alg:KS}.

\begin{algorithm}[!h]
\begin{small}
 		%\SetAlgoNoEnd
		\SetAlgoVlined
 		%\SetAlgoLined
 		\KwData{\textcolor{black}{$(x^{RF^*},y^{RF^*},z^{RF^*})$, $(x^{FO},y^{FO},z^{FO})$, \textsf{AvailTime-RFO}, \textsf{TimeLimit-KS}.}}
 		\KwResult{ A solution to the problem or a warning message.}
        \tcc{Initialization phase}
        \textsf{AvailTime-KS} $\leftarrow$ \textsf{TimeLimit-KS} $+$ \textsf{AvailTime-RFO};\\
        $(x^{*},y^{*},z^{*}) \leftarrow (x^{FO},y^{FO},z^{FO})$; \\
        Define set $\Psi$; \\
        Solve the relaxed problem $(LP)$;\\
        Build kernel $\Delta = (XZ,K)$ and sequence of buckets $B_v$, with $v\in\{1,\ldots,nb\}$;\\
        \textsf{MIP-KS} $\leftarrow$ \textsf{AvailTime-KS}$/3$; \\
        $(x^{(M2)},y^{(M2)},z^{(M2)}) \leftarrow$ solution of the problem $(M2)$ found within the time limit of \textsf{MIP-KS};\\
        \If{$(x^{(M2)},y^{(M2)},z^{(M2)})$ is feasible}{
            $(x^{*},y^{*},z^{*}) \leftarrow (x^{(M2)},y^{(M2)},z^{(M2)})$; \\
        }
        \tcc{Extension phase}
        \For{$v = 1$ \KwTo $nb$}{
            Update \textsf{AvailTime-KS}; \\
            \textsf{MIP-KS} $\leftarrow$ \textsf{AvailTime-KS}$/(nb-v+1)$; \\
            $(x^{(M3)^{v}},y^{(M3)^{v}},z^{(M3)^{v}}) \leftarrow$ solution of the subproblem $(M3)^{v}$ solved within the time limit of \textsf{MIP-KS};\\
            \If{$(x^{(M3)^{v}},y^{(M3)^{v}},z^{(M3)^{v}})$ is feasible}{
                $(x^{*},y^{*},z^{*}) \leftarrow (x^{(M3)^{v}},y^{(M3)^{v}},z^{(M3)^{v}})$; \\
                Update subsets $K$ and $B_{v}$; \\
            }
        }
        \eIf{ $(x^{*},y^{*},z^{*})$ is feasible}{
	        Return $(x^{*},y^{*},z^{*})$. \\
	    }{
            Return warning message stating that the method has not found a feasible solution to ILSSP-NT on parallel machines.
		}
 		\caption{RFO-KS heuristic}\label{alg:KS}
\end{small}
\end{algorithm}

\section{Computational experiments}
\label{TestPre}

This section presents the computational experiments carried out for the comparative analysis with the proposed heuristics. Since the only strategies in the revised literature to specifically solve the ILSSP-NT on parallel machines were tested to solve a small set of problems, we performed two experiments to verify the quality of the solutions obtained by the RFO, RFO-PR and RFO-KS heuristics. In the first experiment, we test the solution methods on the ILSSP-NT on a single machine as in \citep{Guimaraes2013}. In the second experiment, the heuristic methods were used to solve the instance set introduced in \citep{James2011} adapted in this paper to the MIP of the ILSSP-NT on parallel machines and their solutions were analyzed. The results obtained by the proposed heuristics \textcolor{red}{in each experiment} are compared with those found by CPLEX v. 12.10 solver, time limited in 3600 seconds \textcolor{red}{and by RFO-PR* and RFO-KS*, which are variants of RFO-PR and RFO-KS, respectively. Different from RFO-PR and RFO-KS, RFO-PR* and RFO-KS* do not consider the adjustment in the methods to consider the shortcut item characteristics found in the ILSSP-NT. In other words, RFO-PR* and RFO-KS*  define $\Psi$ as an empty set in RFO-PR and RFO-KS, respectively. The main idea of using RFO-PR* and RFO-KS* in the experiments is to demonstrate the relevance in considering the shortcut items in designing the RFO-PR and RFO-KS heuristics to solve the ILSSP-NT.}

The computational tests were carried out in a cluster with 104 computer nodes with two Intel Xeon E5-2680v2 tencore processor of 2.8 GHz, 128 GB DDR3 RAM each. To run the RFO, \textcolor{red}{RFO-PR*, RFO-KS*,} RFO-PR and RFO-KS heuristics, and CPLEX v. 12.10, we used 1 thread and imposed a time limit of 3600 seconds to solve each instance. All algorithms are implemented in C++ language. 

\subsection{Sets of instances}

In the first experiment, we used the 16 classes of 10 instances that \citet{Guimaraes2013} suggested to approach the ILSSP on a single machine with non-triangular setup costs and times. The 160 instances in question have 15 and 25 items ($n$); 10 and 15 periods ($p$); machine capacity usage ($Cut$) of 80\% and 60\%; and setup costs proportional to setup times by a factor ($\Theta$) of 50 and 100. In this sense, each class of instances is referred to as Data$m$-$n$-$p$-$Cut$-$\Theta$.
 
To the second experiment, we adapted the 10 classes of 10 instances initially introduced by \citet{James2011} to approach the ILSSP on parallel machines with capacity variation. The modification was necessary to ensure non-triangular setup costs and times in the 100 instances. The instances used have 2 and 3 parallel machines ($m$); 15 and 20 items ($n$); 5 and 10 periods ($p$); machine capacity usage ($Cut$) of 80\% and 60\%; capacity variation ($CutVar$) of 50\%; setup costs proportional to setup times by a factor ($\Theta$) of 50 and 100; probability of 80\% and 60\% of an extra machine being able to produce a given item ($MProb$); and maximum difference in percentage  of the number of items that could be processed in different machines ($MBal$) of 10\% and 20\%. Accordingly, each class of instances is referred to as Data$m$-$n$-$p$-$Cut$-$\Theta$-$MProb$-$MBal$. Tables \ref{tab:1.1} and \ref{tab:1.2} also show how the parameters were defined to the ILSSP-NT on a single machine and parallel machines according to \citet{James2011} and \citet{Guimaraes2013}.

\begin{table}[!h]
\renewcommand{\arraystretch}{1.5}
\centering
\begin{scriptsize}
\begin{threeparttable}[b]
\caption{Parameter definitions shared by all the 26 classes of instances designed for the ILSSP-NT on a single and parallel machines. Each row presents the problem and the uniform distribution (U$[a,b]$) or value used to define their parameters.}
\label{tab:1.1}
\begin{tabular}{llllll}
\hline
Problem & $b_{ijt}$ & $b{\tnote{*}}_{ijt}$  &$h_{it}$ &$d_{it}$ & $f_{ikt}$  \\
\hline
ILSSP-NT on a single and parallel machines  & U$[5,10]$ & U$[2,4]$ & U$[2,9]$  & U$[40,59]$     & 1   \\
\hline
\end{tabular}
\begin{tablenotes}
     \item[*] Setup times associated to the items (shortcut items) with cleansing properties that incur in cheaper and faster setup costs and times, which can lead to the violation of the triangular inequalities.
   \end{tablenotes}
  \end{threeparttable}
\end{scriptsize}
\end{table}

\begin{table}[!h]
\renewcommand{\arraystretch}{2.0}
\centering
\begin{scriptsize}
\caption{Parameter definitions of the 26 classes of instances designed for the ILSSP-NT on a single and parallel machines. Each row presents a class of instances and the uniform distribution (U$[a,b]$) or value used to define their parameters.}
\label{tab:1.2}
\begin{tabular}{l|l|lccl|l|l}
\cline{1-3}\cline{6-8}
\multicolumn{3}{c}{ILSSP-NT on a single machine}  &  & \multicolumn{1}{l}{} & \multicolumn{3}{c}{ILSSP-NT on parallel machines} \\
\cline{1-3}\cline{6-8}
\multicolumn{1}{l|}{Class of instances}& $T_{kt}$  & $c_{ijk}$  &  & \multicolumn{1}{l}{} & \multicolumn{1}{l|}{Class of instances}& $T_{kt}$  & $c_{ijk}$ \\
\cline{1-3}\cline{6-8}
Data1-15-10-0.6-50  & \multirow{4}{*}{$\left(\mathop {\sum }\limits_{i}d_{it}\times 0.6\right)$}  & \multirow{8}{*}{$(b_{ijt}\times 50)$} &  &   & Data2-15-5-0.8-50-80-20  & \multirow{4}{*}{$\left(\mathop {\sum }\limits_{i}d_{it}\times 0.8\right)$}  & \multirow{4}{*}{$(b_{ijt}\times 50)$}  \\
Data1-15-15-0.6-50  & & & & & Data2-15-10-0.8-50-80-20 &   & \\
Data1-25-10-0.6-50  & & & & & Data3-15-5-0.8-50-80-20  &  &  \\
Data1-25-15-0.6-50  & & &  & & Data3-15-10-0.8-50-80-20 &  &  \\
\cline{1-2}\cline{6-8}
Data1-15-10-0.8-50  & \multirow{4}{*}{$\left(\mathop {\sum }\limits_{i}d_{it}\times 0.8\right)$} &  &  &  & Data3-15-10-0.6-100-60-20  & \multirow{1}{*}{$\left(\mathop {\sum }\limits_{i}d_{it}\times 0.6\right)$} & \multirow{6}{*}{$(b_{ijt}\times 100)$} \\
\cline{6-7}
Data1-15-15-0.8-50  & &  &  & & Data2-15-10-0.8-100-80-20  & \multirow{5}{*}{$\left(\mathop {\sum }\limits_{i}d_{it}\times 0.8\right)$} &  \\
Data1-25-10-0.8-50  &  &  &  &  & Data2-20-10-0.8-100-80-20  &  &  \\
Data1-25-15-0.8-50  &   &    &  &  & Data2-15-10-0.8-100-60-20  &  &  \\
\cline{1-3}
Data1-15-10-0.6-100  & \multirow{4}{*}{$\left(\mathop {\sum }\limits_{i}d_{it}\times 0.6\right)$} & \multirow{8}{*}{$(b_{ijt}\times 100)$} &  &  & Data2-15-10-0.8-100-80-10   &   &  \\
Data1-15-15-0.6-100  &  &  &  &  & Data3-15-10-0.8-100-60-20  &  &  \\
\cline{6-8}
Data1-25-10-0.6-100   &    &   &  & \multicolumn{1}{c}{} & \multicolumn{1}{c}{}  & \multicolumn{1}{c}{} & \multicolumn{1}{c}{}   \\
Data1-25-15-0.6-100  &    &    &  &  \multicolumn{1}{c}{} & \multicolumn{1}{c}{} & \multicolumn{1}{c}{}  & \multicolumn{1}{c}{}   \\
\cline{1-2}
Data1-15-10-0.8-100  & \multirow{4}{*}{$\left(\mathop {\sum }\limits_{i}d_{it}\times 0.8\right)$}  & &  &  \multicolumn{1}{c}{} & \multicolumn{1}{c}{} & \multicolumn{1}{c}{} & \multicolumn{1}{c}{}  \\
Data1-15-15-0.8-100   &  &  &  &   \multicolumn{1}{c}{} & \multicolumn{1}{c}{} & \multicolumn{1}{c}{} & \multicolumn{1}{c}{}  \\
Data1-25-10-0.8-100 & &  &  &   \multicolumn{1}{c}{} & \multicolumn{1}{c}{}  & \multicolumn{1}{c}{} & \multicolumn{1}{c}{}  \\
Data1-25-15-0.8-100  &  &  &  &  \multicolumn{1}{c}{} & \multicolumn{1}{c}{} & \multicolumn{1}{c}{}  & \multicolumn{1}{c}{} \\
\cline{1-3}
\end{tabular}
\end{scriptsize}
\end{table}

To ensure that the instances generated in \citep{James2011} and adapted to tackle the ILSSP-NT on single and parallel machines do not obey the triangular inequality, only the values of the setup costs ($c_{ijk}$) and times ($b_{ijt}$) were modified by inserting shortcut items. Therefore, for each machine of every instance we randomly chose $\lceil n/10 \rceil$ shortcut items and the setup times associated to these items were modified to follow the uniform distribution U$[2,4]$. The remainder setup times preserved the same values defined for the instances of the ILSSP on parallel machines with uniform distribution U$[5,10]$. To maintain the proportionality of the setup costs with the setup times, the setup costs $c_{ijk}$ associated to the shortcut items were modified to ensure the proportionality relation presented in Table \ref{tab:1.2}. Moreover, to guarantee the appropriate cleansing of the machines when producing shortcut items, a minimum of 25 units per production lot was imposed.

\subsection{Evaluation metric}

\textcolor{black}{In the experiments, we assess the optimality gap of CPLEX  v.12.10 time limited in 3600 seconds and the gap between the upper bounds found by the introduced heuristics and the lower bounds found by CPLEX. For this, let $gap_{L}$ be the gap distance between lower bound $Z_{LB}$  provided  by CPLEX v.12.10 in 3600 seconds and a given upper bound $Z_{UB}$. The formula of $gap_{L}$ is presented in  Equation \eqref{gapT}.} 

\begin{equation}
gap_{L} = 100 \times \frac{Z_{UB} - Z_{LB}}{Z_{LB}} \; \% \label{gapT}
\end{equation}

\subsection{Parameter tuning}\label{sec:tuning}

In the attempt to reach the best results with the RFO, RFO-PR and RFO-KS heuristics, the fine tuning of parameters $\lambda$ and $\gamma$ of \textcolor{red}{the RF and FO heuristics} was performed by testing combinations of the values 1, 2, 3, 4 and 5 for all the parameters. \textcolor{red}{In  experiments, we observed that the proposed heuristics provided better results on average to solve instances of the ILSSP-NT when  $\lambda=4$, $\gamma=3$, in the RF heuristic, and  $\lambda = 4$ and $\gamma = 2$, in the FO heuristic.} 

\textcolor{red}{We also tuned $\zeta$, which is a parameter of RFO-PR. The tested values for $\zeta$ were  $p/2$, $2p/3$ and $p$. The results of the experiments indicated that the solutions obtained by RFO-PR are better when $\zeta$ is equal to $2p/3$.} Therefore, the subsequent tests and experiments were carried out considering these parameter values in all heuristics.

All computational experiments were performed considering \textsf{TimeLimit-RF}, \textsf{TimeLimit-FO}, \textsf{TimeLimit-KS} and \textsf{TimeLimit-PR} equal to 1200 seconds. Therefore, the maximum amount of time for the RFO heuristic to obtain a solution was 2400 seconds while for the RFO-PR*, RFO-KS*, RFO-KS and RFO-PR heuristics was 3600 seconds.

\subsection{Experiment I}

To validate the RFO, RFO-PR and RFO-KS heuristics, such heuristics were adapted to solve the ILSSP-NT on a single machine and their results were compared to those obtained by the CPLEX v. \textcolor{black}{12.10} solver \textcolor{red}{and by the RFO-PR* and RFO-KS* heuristics also adapted to solve the ILSSP-NT on a single machine}. We do not compare our results with those obtained by the solution methods reported in \citep{Guimaraes2013} because of the significant differences between the computer configurations used in the experiments, and the CPLEX versions implemented in the matheuristics. Table \ref{tab:1.3} presents the average $gap_{L}$ per class of instances, referred to as $AG_{L}$, as well as the average time (AT) in seconds required by CPLEX, RFO, RFO-PR*, RFO-KS*, RFO-PR and RFO-KS.

\begin{table}[!h]
\renewcommand{\arraystretch}{1.3}
\centering
\begin{scriptsize}
\caption{Results obtained by CPLEX v. 12.10, RFO, RFO-PR*, RFO-KS*, RFO-PR and RFO-KS to the ILSSP-NT on a single machine. Each row reports the average $gap_L$ ($AG_{L}$) of the upper bounds found by the methods and the average time (AT) to solve each  class of instances. }
\label{tab:1.3}
\begin{tabular}{lrrrrrrrrrrrr}
\cline{2-13}
 & \multicolumn{2}{c}{CPLEX} & \multicolumn{2}{c}{RFO} & \multicolumn{2}{c}{RFO-PR*} & \multicolumn{2}{c}{RFO-KS*} & \multicolumn{2}{c}{RFO-PR} & \multicolumn{2}{c}{RFO-KS} \\ \hline
Class of instances & \multicolumn{1}{c}{$AG_{L}$} & \multicolumn{1}{c}{AT} & \multicolumn{1}{c}{$AG_{L}$} & \multicolumn{1}{c}{AT} & \multicolumn{1}{c}{$AG_{L}$} & \multicolumn{1}{c}{AT} & \multicolumn{1}{c}{$AG_{L}$} & \multicolumn{1}{c}{AT} & \multicolumn{1}{c}{$AG_{L}$} & \multicolumn{1}{c}{AT} & \multicolumn{1}{c}{$AG_{L}$} & \multicolumn{1}{c}{AT} \\ \hline
Data1-15-10-0.6-50 & \textbf{0.00} & 7 & 0.01 & 15 & 0.01 & 20 & 0.01 & 21 & 0.01 & 22 & 0.01 & 24 \\
Data1-15-10-0.6-100 & \textbf{0.00} & 361 & 0.10 & 54 & 0.08 & 784 & 0.09 & 184 & 0.06 & 1177 & 0.02 & 709 \\
Data1-15-10-0.8-50 & \textbf{0.00} & 77 & 0.04 & 32 & 0.04 & 42 & 0.03 & 42 & 0.03 & 407 & \textbf{0.01} & 63 \\
Data1-15-10-0.8-100 & \textbf{0.01} & 1615 & 0.20 & 181 & 0.20 & 245 & 0.17 & 813 & 0.16 & 1471 & 0.08 & 1775 \\
Data1-15-15-0.6-50 & \textbf{0.00} & 78 & 0.06 & 37 & 0.06 & 48 & 0.06 & 55 & 0.05 & 410 & 0.04 & 91 \\
Data1-15-15-0.6-100 & \textbf{0.34} & 3343 & 0.43 & 222 & 0.43 & 333 & 0.41 & 1722 & 0.40 & 2361 & 0.40 & 2871 \\
Data1-15-15-0.8-50 & \textbf{0.02} & 1082 & 0.05 & 91 & 0.05 & 472 & 0.05 & 239 & 0.05 & 516 & 0.05 & 599 \\
Data1-15-15-0.8-100 & 0.87 & 3600 & \textbf{0.83} & 208 & \textbf{0.83} & 324 & \textbf{0.83} & 2090 & \textbf{0.83} & 2090 & \textbf{0.83} & 3599 \\
Data1-25-10-0.6-50 & \textbf{0.00} & 97 & 0.04 & 70 & 0.04 & 95 & 0.04 & 99 & 0.03 & 455 & 0.02 & 132 \\
Data1-25-10-0.6-100 & \textbf{0.14} & 2980 & 0.22 & 377 & 0.21 & 815 & 0.19 & 2109 & 0.17 & 2311 & 0.17 & 2750 \\
Data1-25-10-0.8-50 & \textbf{0.00} & 128 & 0.03 & 105 & 0.03 & 130 & 0.03 & 178 & 0.03 & 154 & 0.03 & 397 \\
Data1-25-10-0.8-100 & \textbf{0.44} & 3600 & 0.46 & 1388 & \textbf{0.44} & 2015 & 0.46 & 3100 & 0.45 & 2765 & 0.45 & 3357 \\
Data1-25-15-0.6-50 & \textbf{0.01} & 961 & 0.03 & 160 & 0.03 & 550 & 0.03 & 206 & 0.02 & 900 & 0.03 & 343 \\
Data1-25-15-0.6-100 & 0.70 & 3600 & 0.58 & 742 & 0.57 & 1814 & 0.58 & 3585 & \textbf{0.54} & 3406 & 0.58 & 3600 \\
Data1-25-15-0.8-50 & \textbf{0.01} & 1815 & 0.05 & 490 & 0.05 & 1496 & 0.05 & 811 & 0.05 & 2233 & 0.05 & 1452 \\
Data1-25-15-0.8-100 & 1.21 & 3600 & 0.75 & 1715 & 0.75 & 2820 & 0.75 & 3582 & \textbf{0.74} & 3221 & 0.75 & 3600 \\ \hline
Average & 0.23 & 1684 & 0.24 & 368 & 0.24 & 750 & 0.24 & 1177 & 0.23 & 1494 & \textbf{0.22} & 1585 \\ \hline
\end{tabular}
\end{scriptsize}
\end{table}

One can observe in Table \ref{tab:1.3} that, on average, the RFO, \textcolor{red}{RFO-PR*, RFO-KS*,} RFO-PR and RFO-KS heuristics obtained competitive results to those found by CPLEX, which proved the optimality of 6 classes of instances. RFO-KS had the best mean $AG_L$, followed by RFO-PR\textcolor{red}{, CPLEX, RFO-PR*, RFO-KS* and RFO}. In particular, \textcolor{red}{all the} solution methods achieved upper bounds significantly better than those found by CPLEX in the classes of instances Data1-25-15-0.6-100 and Data1-25-15-0.8-100. \textcolor{red}{RFO-PR* and RFO-KS* had the lowest improvement rates to the initial solutions provided by the RFO heuristic, showing a better performance of the RFO-PR and RFO-KS heuristics.}

Table \ref{tab:1.4} shows the average results of this experiment considering the instances divided into pairs of disjoints groups according to their main characteristics. Each row of this table indicates the average $gap_L$ per group of instances ($AG_L$) that share the following characteristics: number of periods ($p$), number of items ($n$), capacity usage ($Cut$) and proportionality factor between setup costs and times ($\Theta$). The $AG_L$ values marked with a `$\dag$' sign indicate that the results highlighted in bold in the corresponding rows are statistically better than the marked gaps. To assess the statistical significance of the results, we conducted a pairwise t-test~\cite{student1908probable}. We ran this test with the standard significance level of $5\%$. 

\begin{table}[!h]
\renewcommand{\arraystretch}{1.3}
\centering
\begin{scriptsize}
\caption{Results obtained by CPLEX v. 12.10, RFO, RFO-PR*, RFO-KS*, RFO-PR and RFO-KS per group of instances combined according to their main characteristics:  $p$ (15 and 25), $n$ (10 and 15), $Cut$ (0.6 and 0.8) and  $\Theta$ (50 and 100). Each row reports the average $gap_L$ ($AG_{L}$) of the upper bounds obtained by the methods to solve each group of instances.}
\label{tab:1.4}
\begin{tabular}{lllrrrrrrrrrrr}
\cline{4-14}
 &  &  & \multicolumn{1}{l}{CPLEX} & \multicolumn{1}{l}{} & \multicolumn{1}{l}{RFO} & \multicolumn{1}{l}{} & \multicolumn{1}{l}{RFO-PR*} & \multicolumn{1}{l}{} & \multicolumn{1}{l}{RFO-KS*} & \multicolumn{1}{l}{} & \multicolumn{1}{l}{RFO-PR} & \multicolumn{1}{l}{} & \multicolumn{1}{l}{RFO-KS} \\ \cline{1-2} \cline{4-4} \cline{6-6} \cline{8-8} \cline{10-10} \cline{12-12} \cline{14-14} 
Parameter & Values &  & \multicolumn{1}{c}{$AG_{L}$} & \multicolumn{1}{l}{} & \multicolumn{1}{c}{$AG_{L}$} & \multicolumn{1}{l}{} & \multicolumn{1}{c}{$AG_{L}$} & \multicolumn{1}{c}{} & \multicolumn{1}{c}{$AG_{L}$} & \multicolumn{1}{l}{} & \multicolumn{1}{c}{$AG_{L}$} & \multicolumn{1}{l}{} & \multicolumn{1}{c}{$AG_{L}$} \\ \hline
\multirow{2}{*}{$p$} & 15 &  & \textbf{0.155} &  & 0.216$^{\dag}$ &  & 0.213$^{\dag}$ &  & 0.208$^{\dag}$ &  & 0.201$^{\dag}$ &  & 0.181$^{\dag}$ \\
 & 25 &  & 0.313$^{\dag}$ &  & 0.269$^{\dag}$ &  & 0.264 &  & 0.267$^{\dag}$ &  & \textbf{0.252} &  & 0.259$^{\dag}$ \\ \hline
\multirow{2}{*}{$n$} & 10 &  & \textbf{0.072} &  & 0.135$^{\dag}$ &  & 0.130$^{\dag}$ &  & 0.127$^{\dag}$ &  & 0.119$^{\dag}$ &  & 0.099$^{\dag}$ \\
 & 15 &  & 0.395$^{\dag}$ &  & 0.350$^{\dag}$ &  & 0.347$^{\dag}$ &  & 0.347$^{\dag}$ &  & \textbf{0.334} &  & 0.342 \\ \hline
\multirow{2}{*}{$Cut$} & 0.6 &  & \textbf{0.148} &  & 0.184$^{\dag}$ &  & 0.179$^{\dag}$ &  & 0.178 &  & 0.160 &  & 0.158 \\
 & 0.8 &  & 0.320 &  & 0.302$^{\dag}$ &  & 0.298 &  & 0.297$^{\dag}$ &  & 0.293 &  & \textbf{0.282} \\ \hline
\multirow{2}{*}{$\Theta$} & 50 &  & \textbf{0.005} &  & 0.040$^{\dag}$ &  & 0.039$^{\dag}$ &  & 0.039$^{\dag}$ &  & 0.034$^{\dag}$ &  & 0.030$^{\dag}$ \\
 & 100 &  & 0.463$^{\dag}$ &  & 0.446$^{\dag}$ &  & 0.438$^{\dag}$ &  & 0.436$^{\dag}$ &  & 0.419 &  & \textbf{0.410} \\ \hline
\end{tabular}
\end{scriptsize}
\end{table}

\textcolor{black}{According to Table \ref{tab:1.4}, RFO-PR outperformed \textcolor{red}{with statistical significance: (i) RFO  in every group of instances; (ii) RFO-PR* in all the groups except for the one with the highest  $Cut$ value; (iii) RFO-KS* in the groups with the highest values of $p$ and $n$,  with $Cut$ equals to 0.6, and in the groups with both values of $\Theta$; and (iv) } CPLEX in the groups with the highest $p$ and $n$  values. RFO-KS performed \textcolor{red}{statistically better than: (i) RFO and RFO-KS* in all the groups; (ii) RFO-PR* in the group with the highest $\Theta$ besides the groups with the lowest parameter values; and (iii)  CPLEX in the groups with the highest values of $p$, $n$ and $\Theta$}. CPLEX outperformed RFO, RFO-PR*, RFO-KS*, RFO-PR and RFO-KS with statistical significance in the groups with the lowest parameter values.}

\begin{figure}[htbp]
\center
\begin{tikzpicture}
\pgfplotsset{every axis legend/.append style={
at={(0.8,0.05)},
anchor=south}
},
\begin{axis}[
width=14cm,
height=7cm,
%y axis line style={opacity=0},
%xmode=0.002,
%log basis x=10,
legend columns=1,
xmin=0.99997,
xmax=1.0051,
xtick={1,1.001,1.002,1.003,1.004,1.005},
xticklabels={1,1.001,1.002,1.003,1.004,1.005},
ymin=0.45,
ymax=1.02,
grid=major,
xlabel={$\tau$},
ylabel={$\Theta(\tau)$ },
]

\addplot[/tikz/solid, blue!50!white,line width=1pt]
table[x=x,y=y] {Tudo-SM-CPLEX.txt};

\addplot[/tikz/solid, red!70!white,line width=1.5pt]
table[x=x,y=y] {Tudo-SM-RFO.txt};

\addplot[/tikz/solid, black!60!white,line width=2pt]
table[x=x,y=y] {Tudo-SM-RFO-PR1.txt};

\addplot[/tikz/densely dashed, green!70!black,line width=2pt]
table[x=x,y=y] {Tudo-SM-RFO-KS1.txt};

\addplot[/tikz/dotted, orange!70!white,line width=3pt]
table[x=x,y=y] {Tudo-SM-RFO-PR2.txt};

\addplot[/tikz/dashdotdotted, violet!70!white,line width=2.5pt]
table[x=x,y=y] {Tudo-SM-RFO-KS2.txt};

\legend{CPLEX, RFO, RFO-PR*, RFO-KS*, RFO-PR, RFO-KS}
\end{axis}
\end{tikzpicture}
\caption{Performance profiles associated to the upper bounds obtained by CPLEX, RFO, RFO-PR*, RFO-KS*, RFO-PR and RFO-KS to solve the ILSSP-NT on a single machine.}
\label{fig:dolanexp1}
\end{figure}
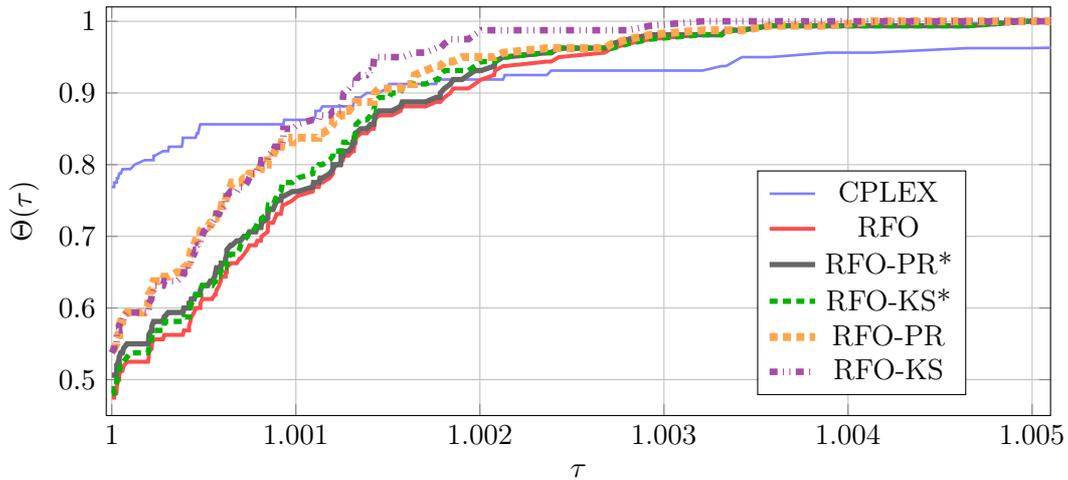

\textcolor{black}{In Figure~\ref{fig:dolanexp1}, we present the performance profiles introduced in \cite{Dolan2002} contrasting the upper bounds found by the \textcolor{red}{six} methods. In this evaluation, the performance profile of a method consists of the cumulative distribution function for the objective function. It is possible to observe that CPLEX achieved the best upper bounds in 76\% of the instances, whereas RFO-PR and RFO-KS obtained the best solutions in approximately 54\% of the tested instances. RFO found the best solutions in around 47\% of the instances, \textcolor{red}{followed by RFO-KS* and RFO-PR* with the best solutions in approximately 48\% and 50\% of the instances, respectively}. It is possible to observe in the graphic that RFO-KS had the best results in almost all instances when $\tau$ is just 1.002. This means that in almost all the 46\% of instances that such method did not obtain the best upper bounds, its values were worse than the best by a factor of at most 0.2\%. Besides, the curve of CPLEX crossed with the profiles of \textcolor{red}{RFO-KS* and} RFO-PR for $\tau$ equals \textcolor{red}{1.0015}. This means that the \textcolor{red}{three} solution methods have the same percentage of solutions whose quality is at most 0.15\% worst than the best overall solutions. Moreover, according to the performance profiles, \textcolor{red}{RFO-KS dominates CPLEX and the other methods} for $\tau\geq 1.0013$.}

\textcolor{black}{Therefore, the results of this experiment indicate that the introduced methods, in particular, RFO-KS, have a good performance in solving ILSSP-NT on a single machine. The next section presents the results of the introduced methods to the ILSSP-NT on parallel machines.}

\subsection{Experiment II}

This section is dedicated to attest the good performance of the RFO, RFO-PR and RFO-KS heuristics proposed to solve ILSSP-NT on parallel machines. Table \ref{tab:2.1} presents the average $gap_{L}$, denoted by $AG_{L}$, obtained by the methods and the respective average times (AT) in seconds.

\begin{table}[!h]
\renewcommand{\arraystretch}{1.3}
\centering
\begin{scriptsize}
\caption{Results obtained by CPLEX v. 12.10, RFO, RFO-PR*, RFO-KS*, RFO-PR and RFO-KS to the ILSSP-NT on parallel machines. Each row reports the average $gap_L$ ($AG_{L}$) of the upper bounds found by the methods and the average time (AT) in seconds to solve each class of instances.}
\label{tab:2.1}
\begin{tabular}{llrrrrrrrrrrr}
\cline{2-13}
 & \multicolumn{2}{c}{CPLEX} & \multicolumn{2}{c}{RFO} & \multicolumn{2}{c}{RFO-PR*} & \multicolumn{2}{c}{RFO-KS*} & \multicolumn{2}{c}{RFO-PR} & \multicolumn{2}{c}{RFO-KS} \\ \hline
Class of instances & $AG_{L}$ & AT & $AG_{L}$ & AT & $AG_{L}$ & AT & $AG_{L}$ & AT & $AG_{L}$ & AT & $AG_{L}$ & AT \\ \hline
Data2-15-5-0.8-50-80-20 & \textbf{0.00} & 24 & 0.08 & 36 & 0.08 & 37 & 0.08 & 38 & 0.08 & 52 & 0.08 & 39 \\
Data2-15-10-0.8-50-80-20 & 0.81 & 3504 & 0.82 & 51 & 0.77 & 718 & 0.81 & 52 & \textbf{0.68} & 1857 & 0.81 & 676 \\
Data2-15-10-0.8-100-60-20 & \textbf{1.32} & 3289 & 1.40 & 62 & 1.34 & 873 & 1.40 & 713 & 1.33 & 2133 & 1.36 & 1243 \\
Data2-15-10-0.8-100-80-10 & 4.66 & 3600 & 3.76 & 2125 & 3.70 & 3062 & 3.70 & 2426 & 3.51 & 2816 & \textbf{3.48} & 2222 \\
Data2-15-10-0.8-100-80-20 & 3.70 & 3600 & 3.50 & 1951 & 3.46 & 2936 & 3.46 & 2085 & \textbf{3.16} & 2745 & \textbf{3.16} & 1998 \\
Data2-20-10-0.8-100-80-20 & 2.45 & 3600 & 2.28 & 2271 & 2.26 & 3393 & 2.26 & 2594 & 2.08 & 3282 & \textbf{2.08} & 2496 \\
Data3-15-5-0.8-50-80-20 & \textbf{0.31} & 1475 & 0.46 & 780 & 0.43 & 1197 & 0.45 & 788 & 0.40 & 1738 & 0.45 & 897 \\
Data3-15-10-0.6-100-60-20 & 1.69 & 3275 & 1.70 & 841 & 1.67 & 1068 & 1.70 & 860 & \textbf{1.54} & 2819 & 1.66 & 1512 \\
Data3-15-10-0.8-50-80-20 & 2.30 & 3600 & 2.17 & 1592 & 1.95 & 2388 & 2.16 & 1621 & \textbf{1.82} & 2411 & 2.01 & 1645 \\
Data3-15-10-0.8-100-60-20 & 4.79 & 3600 & 5.07 & 1954 & 4.94 & 2995 & 5.02 & 2013 & \textbf{4.50} & 3007 & 4.55 & 2021 \\ \hline
Average & 2.20 & 2957 & 2.12 & 1269 & 2.06 & 1867 & 2.10 & 1367 & \textbf{1.91} & 2286 & 1.96 & 1475 \\ \hline
\end{tabular}
\end{scriptsize}
\end{table}

Table~\ref{tab:2.1} shows that RFO, RFO-PR*, RFO-KS*, RFO-PR and RFO-KS obtained very competitive results to those found by CPLEX. Both RFO-PR and RFO-KS achieved the best overall averages  \textcolor{red}{1.91\% and 1.96\%, respectively, in contrast to} the average of 2.20\% obtained by CPLEX. \textcolor{red}{Regarding} the mean running times, RFO-KS was significantly faster than RFO-PR and CPLEX. \textcolor{red}{In comparison to the solutions obtained by RFO, RFO-KS* provided the worst improvement rates, on average,  followed by RFO-PR*.}

A second analysis of the results considering the instances divided into pairs of disjoint groups in 7 different ways according to the parameters  $m$, $p$, $n$, $Cut$, $\Theta$, $MProb$ and $MBal$ is presented in Table \ref{tab:2.2}. Each row of Table \ref{tab:2.2} indicates the average $gap_L$ ($AG_L$) per group of instances that share the same parameter values indicated in the first columns. The $AG_L$ values marked with a `$\dag$' sign indicate that the results highlighted in bold in the corresponding rows are statistically better than the marked gaps. Again, to assess the statistical significance of the results, we conducted a pairwise t-test  ~\cite{student1908probable}. We ran this test with the standard significance level of $5\%$.

\begin{table}[!h]
\renewcommand{\arraystretch}{1.3}
\centering
\begin{scriptsize}
\caption{Results obtained by CPLEX v. 12.10, RFO, RFO-PR*, RFO-KS*, RFO-PR and RFO-KS per group of instances combined according to their main characteristics:  $m$ (2 and 3), $p$ (15 and 20), $n$ (5 and 10), $Cut$ (0.6 and 0.8),  $\Theta$ (50 and 100), $MProb$ (60 and 80) and $MBal$ (10 and 20). Each row reports the average $gap_L$ ($AG_{L}$) of the upper bounds found by the methods after solving each group of instances.}
\label{tab:2.2}
\begin{tabular}{cclrrrrrrrrrrr}
\cline{4-14}
 &     &       &
  \multicolumn{1}{c}{CPLEX} &  \multicolumn{1}{c}{} &  \multicolumn{1}{c}{RFO} &
  \multicolumn{1}{c}{} &  \multicolumn{1}{c}{RFO-PR*} &  \multicolumn{1}{c}{} &
  \multicolumn{1}{c}{RFO-KS*} &  \multicolumn{1}{c}{} &  \multicolumn{1}{c}{RFO-PR} &
  \multicolumn{1}{c}{} &
  \multicolumn{1}{c}{RFO-KS} \\ \cline{1-2} \cline{4-4} \cline{6-6} \cline{8-8} \cline{10-10} \cline{12-12} \cline{14-14} 
Parameter &   Value &      &
  \multicolumn{1}{c}{$AG_{L}$} &  \multicolumn{1}{c}{} &  \multicolumn{1}{c}{$AG_{L}$} &
  \multicolumn{1}{c}{} &  \multicolumn{1}{c}{$AG_{L}$} &  \multicolumn{1}{c}{} &
  \multicolumn{1}{c}{$AG_{L}$} &  \multicolumn{1}{c}{} &  \multicolumn{1}{c}{$AG_{L}$} &
  \multicolumn{1}{c}{} &  \multicolumn{1}{c}{$AG_{L}$} \\ \hline
\multirow{2}{*}{$m$}        & 2   &  & 2.139$^{\dag}$ &  & 1.954$^{\dag}$ &  & 1.918$^{\dag}$ &  & 1.936$^{\dag}$ &  & \textbf{1.790} &  & 1.814          \\
                          & 3   &  & 2.276$^{\dag}$         &  & 2.351$^{\dag}$          &  & 2.252$^{\dag}$          &  & 2.337$^{\dag}$          &  & \textbf{2.070} &  & 2.173$^{\dag}$ \\ \hline
\multirow{2}{*}{$p$}        & 15  &  & 2.204$^{\dag}$ &  & 2.129$^{\dag}$ &  & 2.062$^{\dag}$ &  & 2.113$^{\dag}$ &  & \textbf{1.915} &  & 1.976$^{\dag}$ \\
                          & 20  &  & 2.451$^{\dag}$ &  & 2.282          &  & 2.269          &  & 2.266          &  & 2.085          &  & \textbf{2.083} \\ \hline
\multirow{2}{*}{$n$}        & 5   &  & \textbf{0.166} &  & 0.280$^{\dag}$ &  & 0.265$^{\dag}$ &  & 0.280$^{\dag}$ &  & 0.250$^{\dag}$ &  & 0.280$^{\dag}$ \\
                          & 10  &  & 2.719$^{\dag}$ &  & 2.588$^{\dag}$ &  & 2.515$^{\dag}$ &  & 2.567$^{\dag}$ &  & \textbf{2.331} &  & 2.393$^{\dag}$ \\ \hline
\multirow{2}{*}{$Cut$}      & 0.6 &  & 1.696 &  & 1.700$^{\dag}$ &  & 1.674$^{\dag}$ &  & 1.700$^{\dag}$ &  & \textbf{1.546} &  & 1.665          \\
                          & 0.8 &  & 2.289$^{\dag}$ &  & 2.195$^{\dag}$ &  & 2.129$^{\dag}$ &  & 2.176$^{\dag}$         &  & \textbf{1.975} &  & 2.023$^{\dag}$ \\ \hline
\multirow{2}{*}{$\Theta$} & 50  &  & 0.880 &  & 0.902 &  & 0.828 &  & 0.902 &  & \textbf{0.764} &  & 0.862          \\
                          & 100 &  & 3.106$^{\dag}$ &  & 2.953$^{\dag}$ &  & 2.899$^{\dag}$ &  & 2.925$^{\dag}$ &  & \textbf{2.691} &  & 2.719$^{\dag}$          \\ \hline
\multirow{2}{*}{$Mprob$}    & 60  &  & 2.603 &  & 2.725$^{\dag}$ &  & 2.655$^{\dag}$ &  & 2.707$^{\dag}$ &  & \textbf{2.462} &  & 2.527$^{\dag}$ \\
                          & 80  &  & 2.067$^{\dag}$ &  & 1.893$^{\dag}$ &  & 1.834$^{\dag}$ &  & 1.877$^{\dag}$ &  & \textbf{1.702} &  & 1.752          \\ \hline
\multirow{2}{*}{$Mbal$}     & 10  &  & 4.668$^{\dag}$ &  & 3.762 &  & 3.700 &  & 3.700 &  & 3.515 &  & \textbf{3.487} \\
                          & 20  &  & 1.955$^{\dag}$ &  & 1.963$^{\dag}$ &  & 1.902$^{\dag}$ &  & 1.952$^{\dag}$ &  & \textbf{1.754} &  & 1.819$^{\dag}$ \\ \hline
\end{tabular}
\end{scriptsize}
\end{table}

According to Table \ref{tab:2.2}, both RFO-PR and RFO-KS outperformed CPLEX with statistical significance in at least one of the groups per parameter. In particular, when the instances were grouped considering parameters $p$ and $MBal$, RFO-PR and RFO-KS outperformed CPLEX in both groups. \textcolor{red}{RFO-PR* and RFO-KS* were better than CPLEX in the groups with the lowest  $m$ and $Mbal$ values, and in the group with the highest  $Mprob$. Moreover, RFO-PR* also outperformed CPLEX in the groups with the highest  $n$ and $Cut$ values.} CPLEX outperformed \textcolor{red}{RFO-PR*, RFO-KS*,} RFO-PR and RFO-KS with statistical significance in only one of the groups of instances\textcolor{red}{, and RFO in two groups}. These results show that RFO-PR and  RFO-KS excelled at solving the ILSSP-NT on parallel machines.

\textcolor{red}{Besides, considering the results presented in Table~\ref{tab:2.2}, we carried out a pairwise t-test to assess the impact of the intensification strategies on the quality of the obtained solutions and the superiority of RFO-KS and RFO-PR over the other methods. The results indicated that RFO-PR and RFO-KS had significantly better gaps \textcolor{red}{in all the groups with the highest values per parameter. Apart from the groups with the lowest values of $n$, $\Theta$ and $Mbal$, and with the highest $p$, RFO-PR outperformed RFO, RF-PR* and RFO-KS* in all other groups. RFO-KS was better than RFO and RFO-KS* in the groups with the lowest values of $m$, $p$ and $Mprob$, and highest $n$, $Cut$ and $Mbal$ values. Moreover, RFO-KS outperformed RFO-PR* with statistical significance in the groups with the lowest  $m$, and highest values of $n$, $Cut$, $\Theta$ and $Mbal$.} These results demonstrate that (i) the hybridization of RFO with the intensification strategies provided a significant improvement in the method; and (ii) the consideration of shortcut items in KS and PR provided better results for the hybrid heuristics.}

\begin{figure}[htbp]
\center
\begin{tikzpicture}
\pgfplotsset{every axis legend/.append style={
at={(0.8,0.05)},
anchor=south}
},
\begin{axis}[
width=14cm,
height=7cm,
%y axis line style={opacity=0},
%xmode=0.002,
%log basis x=10,
legend columns=1,
xmin=0.9998,
xmax=1.023,
xtick={1.000,1.005,1.010,1.015,1.020},
xticklabels={1,1.005,1.010,1.015,1.020},
ymin=0.3,
ymax=1.02,
grid=major,
xlabel={$\tau$},
ylabel={$\Theta(\tau)$ },
]

\addplot[/tikz/solid, blue!50!white,line width=1pt]
table[x=x,y=y] {Tudo-PM-CPLEX.txt};

\addplot[/tikz/solid, red!70!white,line width=2pt]
table[x=x,y=y] {Tudo-PM-RFO.txt};

\addplot[/tikz/solid, black!60!white,line width=3pt]
table[x=x,y=y] {Tudo-PM-RFO-PR1.txt};

\addplot[/tikz/densely dashed, green!70!black,line width=2pt]
table[x=x,y=y] {Tudo-PM-RFO-KS1.txt};

\addplot[/tikz/dotted, orange!70!white,line width=3pt]
table[x=x,y=y] {Tudo-PM-RFO-PR2.txt};

\addplot[/tikz/dashdotdotted, violet!70!white,line width=2.5pt]
table[x=x,y=y] {Tudo-PM-RFO-KS2.txt};

\legend{CPLEX, RFO, RFO-PR*, RFO-KS*, RFO-PR, RFO-KS}
\end{axis}
\end{tikzpicture}
\caption{Performance profiles associated to the upper bounds obtained by CPLEX, RFO, RFO-PR*, RFO-KS*, RFO-PR and RFO-KS to solve the ILSSP-NT on parallel machine.}
\label{fig:dolanexp2}
\end{figure}
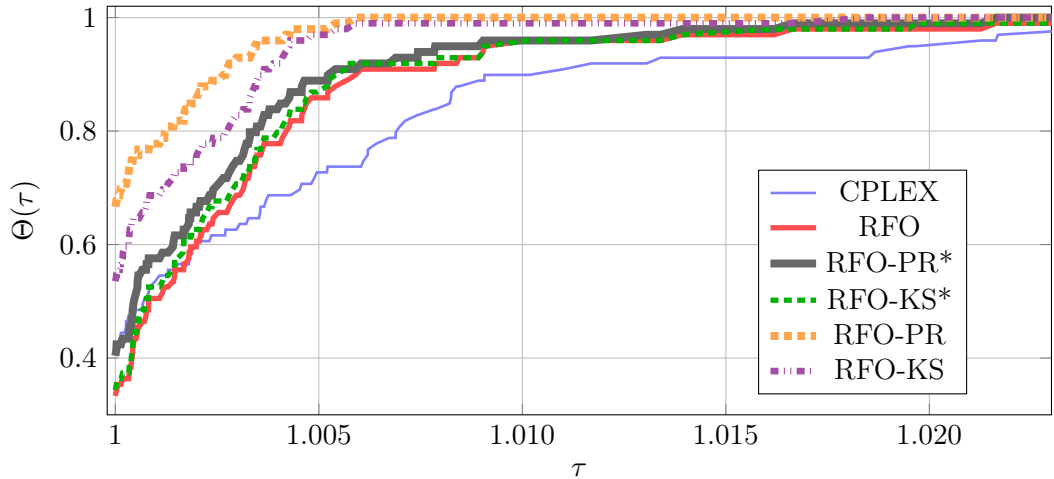

Figure~\ref{fig:dolanexp2} shows the performance profiles introduced in \cite{Dolan2002} contrasting the upper bounds found by the six methods. It is possible to observe that RFO-KS and RFO-PR achieved the best upper bounds in 66\% and 63\% of the instances, respectively, whereas \textcolor{red}{RFO, RFO-PR*, RFO-KS* and CPLEX obtained the best solutions in approximately 33\%, 40\%, 34\% and 42\%}, respectively. It is also possible to observe in the graphic that RFO-PR had the best results in all instances when $\tau$ is just \textcolor{red}{1.006}. Besides, the curves of RFO-KS and RFO-PR dominated the CPLEX\textcolor{red}{, RFO, RFO-PR* and RFO-KS*} curves. These results state the outstanding results of both RFO-KS and RFO-PR to the ILSSP-NT on parallel machines.

\section{Final remarks}
\label{conclu}

This paper tackled the integrated lot sizing and scheduling problem on non-identical parallel machines with capacity constraints, setup carry-over and sequence-dependent setup costs and times that are not-triangular, referred to here as ILSSP-NT on parallel machines. The ILSSP-NT on parallel machines is of utmost importance to the rising sector of food processing industry, for example. It is a very challenging problem and, to the best of our knowledge, \cite{Kang1999} and \cite{Meyr02} were the only studies to address this problem. More attention was given to its variants such as the ILSSP-NT on a single machine and the ILSSP on parallel machines. After a thorough literature review on related problems, we decided to adapt the mathematical formulation proposed in \citep{Guimaraes14} to approach the ILSS-NT on parallel machines.

MIP-based heuristics have been extensively used to solve most of the problems related to the ILSSP-NT on parallel machines. Bearing this in mind, in order to efficiently solve the ILSSP-NT on parallel machines, in this paper we proposed three matheuristics referred to as RFO, RFO-PR and RFO-KS. The RFO heuristic is a one-step method that combines a simple version of the RF and the FO heuristics to find solutions to the ILSSP-NT on parallel machines. The RFO-PR and RFO-KS heuristics are two-step methods: in the first step the RF and the FO heuristics are combined (RFO) to find initial solutions; and in the second step, the intensification-based heuristics path-relinking and kernel search are used to refine the initial solutions in the RFO-PR and RFO-KS, respectively. 

For a thorough comparative analysis, \textcolor{red}{two experiments were performed with} the proposed methods to approach the ILSSP-NT on a single machine and the ILSSP-NT on parallel machines using the instance sets introduced in \citep{Guimaraes2013} and in \citep{James2011}, respectively. In the first experiment, we attest that the RFO, RFO-PR and RFO-KS heuristics can offer good quality solutions to the ILSSP-NT on a single machine as well as the CPLEX v. 12.10 solver time limited in 3600 seconds. The second experiment contrasting the results obtained by the RFO, RFO-PR and RFO-KS heuristics and CPLEX v. 12.10 solver time limited to solve the ILSSP-NT on parallel machines demonstrates the best performance of the proposed heuristics. Moreover, we show that RFO-KS and RFO-PR improved the solutions found by RFO in most of the cases with statistical significance. \textcolor{red}{In addition,  to validate the adjustments specially designed to tackle the ILSSP-NT, we considered the variants of RFO-PR and RFO-KS without the adjustments,  called RFO-PR* and RFO-KS*, respectively. Both experiments provide an analysis that shows the better performance of the proposed  RFO-PR and RFO-KS  over  RFO-PR* and RFO-KS*, respectively.}

\section{Acknowledgments}
\label{agradecimentos}

We are thankful to the financial support from the Coordenação de Aperfeiçoamento de Pessoal de Nível Superior (CAPES), to Fundação de Amparo à Pesquisa do Estado de São Paulo (FAPESP) under grant numbers 16/02537-0 and 15/21660-4, and to Conselho Nacional de Desenvolvimento Científico e Tecnológico (CNPq) under grant number 306036/2018-56. Research carried out using the computational resources of the Center for Mathematical Sciences Applied to Industry (CeMEAI) funded by FAPESP (grant 2013/07375-0). The second author also thanks Leonardo V. Rosset for giving her a hand. We are also very thankful for the valuable comments of the anonymous referees.

\section*{References}
\bibliographystyle{apa}
\bibliography{Doutorado_bibv1}

\end{document}

\appendix
\section{Pseudo-codes of the RF and FO heuristics} \label{appendix}

\textcolor{red}{This section presents the pseudo-codes of RF and FO heuristics in  Algorithms \ref{alg:RF} and \ref{alg:FO}, respectively}.
%the strategy described in Section~\ref{sec:RF} is presented in
\begin{algorithm}[!h]
\begin{small}
 		%\SetAlgoNoEnd
		\SetAlgoVlined
 		%\SetAlgoLined
 		\KwData{Instance, $\lambda$, $\gamma$, \textsf{TimeLimit-RF}.}
 		\KwResult{The solution found in iteration $\theta-3$ and the best solution found to the problem.}    %($x^{RF},y^{RF},z^{RF}$)
 		$S^ {\phi} \leftarrow \{ (i,k,\phi,u): i \in \{1,\ldots,n\}, k \in \{1,\ldots,m\}; u \in \{\phi,\ldots,p\}\}, \quad  \forall \; \phi = 1, \ldots, p$; \\
 		\textcolor{black}{$\theta \leftarrow \lceil(p-\lambda) /(\lambda -\gamma)\rceil+1$;} \\
 		\textcolor{black}{Define $[t'_{v},t^{"}_{v}],\,  \forall \; v = 1, \ldots, \theta$, as described in Step 1 of RF heuristic;}\\
 		\textcolor{black}{\textsf{MIP-RF} $\leftarrow$ \textsf{TimeLimit-RF}$/\theta$;} \\
		$v \leftarrow 1$;\\
        \While{ any stop criteria is not reached}{
		    $(x^{(MIP)^{v}},y^{(MIP)^{v}},z^{(MIP)^{v}})$ $\leftarrow $ solution of problem $(MIP)^{v}$ \textcolor{black}{within the time limit of \textsf{MIP-RF};}\\
			$v \leftarrow v + 1$;\\
	    }
        Return \textcolor{black}{$\{(x^{(MIP)^{\theta -3}},y^{(MIP)^{\theta -3}},z^{(MIP)^{\theta -3}}), (x^{(MIP)^{v}},y^{(MIP)^{v}},z^{(MIP)^{v}})\}$}. \\
 		\caption{RF heuristic}\label{alg:RF}
\end{small}
\end{algorithm}

%\section{Pseudo-code of the FO heuristic}

%A pseudo-code of the FO heuristic presented in Section \ref{sec:FO} is presented in Algorithm \ref{alg:FO}.

\begin{algorithm}[!h]
    \begin{small}
             		%\SetAlgoNoEnd
            		\SetAlgoVlined
             		%\SetAlgoLined
             		\KwData{$(x^{*},y^{*},z^{*})$, $\lambda$, $\gamma$, \textsf{TimeLimit-FO}.}
             		\KwResult{Updated $(x^{*},y^{*},z^{*})$.}
             		$S^ {\phi} \leftarrow \{ (i,k,t,u): i \in \{1,\ldots,n\},\, k \in \{1,\ldots,m\};\, t = \phi;\, u \in \{1,\ldots,p\}\}, \quad  \forall \; \phi = 1, \ldots, p$; \\
             		$\theta \leftarrow \lceil(p-\lambda)/(\lambda -\gamma)\rceil+1$; \\
 					Define period intervals $[t'_{v},t^{"}_{v}],\,  \forall \; v = 1, \ldots, \theta$, as described in Step 1 of the FO heuristic;\\
 					\textsf{MIP-FO} $\leftarrow$ \textsf{TimeLimit-FO}$/\theta$; \\
             		\textsf{ElapsedTime-FO} $\leftarrow$ 0; \\
					\textcolor{black}{$v \leftarrow 1$;}\\
             		\textcolor{black}{$search \leftarrow 0$;} \\
                    \While{\textcolor{black}{\textup{\textsf{ElapsedTime-FO}} $< $ \textup{\textsf{TimeLimit-FO}} and $v \neq 0$}}{
                        \textcolor{black}{$(x^{(MIP)^{v}},y^{(MIP)^{v}},z^{(MIP)^{v}})\leftarrow $ solution of problem $(MIP)^{v}$ within the time limit of \textsf{MIP-FO};} \\
                        \If{\textcolor{black}{$Z^{(MIP)^{v}} < Z^{*}$}}{
                            \textcolor{black}{$(x^{*},y^{*},z^{*}) \leftarrow (x^{(MIP)^{v}},y^{(MIP)^{v}},z^{(MIP)^{v}})$;}\\
                            \textcolor{black}{$search \leftarrow 1$;} \\
                        }
                        \eIf{\textcolor{black}{$v < \theta$}}{
                            \textcolor{black}{$v \leftarrow v + 1$;} \\
                        }
                        {
                        	\eIf{\textcolor{black}{$search == 1$}}{
                        		\textcolor{black}{$v \leftarrow 1$;} \\
                        		\textcolor{black}{$search \leftarrow 0$;} \\
                            }
                            {
                            	\textcolor{black}{$v \leftarrow 0$;} \\
                            }   
                        }
                        \textcolor{black}{Update \textsf{ElapsedTime-FO};} \\
                        \textcolor{black}{\textsf{MIP-FO} $\leftarrow($\textsf{TimeLimit-FO} - \textsf{ElapsedTime-FO}$)/(\theta-v+1)$;} \\
                    }
            	    Return $(x^{*},y^{*},z^{*})$; \\
             		\caption{FO heuristic}\label{alg:FO}
        \end{small}
\end{algorithm}